\magnification 1200
\nopagenumbers
\headline={\hss \tenrm \folio \hss}
\voffset=3\baselineskip
\centerline {\bf Quantum Macrostatistical Theory of 
Nonequilibrium Steady States} 
\vskip 0.5cm 
\centerline {{\bf by Geoffrey L. Sewell}\footnote{$^{a}$}{E-mail 
address g.l.sewell@qmul.ac.uk}} 
\vskip 0.5cm 
\centerline {\bf Department of Physics, Queen Mary, University 
of London} 
\vskip 0.2cm 
\centerline {\bf Mile End Road, London E1 4NS, UK} 
\vskip 1cm\noindent 
{\bf Abstract.} We provide a general macrostatistical formulation of 
nonequilibrium steady states of reservoir driven quantum systems. 
This formulation is centred on the large scale properties of the 
locally conserved hydrodynamical observables, and our basic physical 
assumptions comprise (a) a chaoticity hypothesis for the nonconserved 
currents carried by these observables, (b) an extension of Onsager\rq s 
regression hypothesis to fluctuations about nonequilibrium states, and 
(c) a certain mesoscopic local equilibrium hypothesis. On this basis we 
obtain a picture wherein the fluctuations of the hydrodynamical 
variables about a nonequilibrium steady state execute a Gaussian Markov 
process of a generalized Onsager-Machlup type, which is completely 
determined by the position dependent transport coefficients and the 
equilibrium entropy function of the system. This picture reveals that 
the transport coefficients satisfy a generalized form of the Onsager 
reciprocity relations in the nonequilibrium situation and that the 
spatial correlations of the hydrodynamical observables are generically 
of long range. This last result constitutes a model-independent quantum 
mechanical generalization of that obtained for special classical 
stochastic systems and marks a striking difference between the steady 
nonequilibrium and equilibrium states, since it is only at critical 
points that the latter carry long range correlations. 
\vfill\eject 
\centerline {\bf 1. Introduction} 
\vskip 0.3cm 
The statistical thermodynamics of nonequilibrium steady states or, more 
generally, dynamically stable ones, of reservoir driven macroscopic 
systems\footnote{$^{b}$}{A very simple example of such a state is the 
stationary one of a solid rod, whose ends are coupled to thermostats of 
different temperatures.} is a key area of the natural sciences, with 
ramifications for condensed matter physics [1-4], chemistry [5] and 
biology [6]. At the phenomenological and heuristic levels, there is an 
abundant literature on this subject. At the level of mathematical 
physics, however, the subject is still at an exploratory stage. In the 
classical regime, two types of rigorous approaches have been made to 
it. The first is centred on the hypotheses that the macroscopic 
properties of complex systems are yielded by the model of classical 
Anosov dynamical systems [7,8]. This hypothesis is designed to capture 
the chaoticity that underlies macroscopic irreversibility, and it has 
been shown to lead to nonequilibrium generalizations both of the 
Onsager reciprocity relations [8] and of the fluctuation-dissipation 
theorem [7]. A second approach is centred on microscopic treatments of 
stochastic (non-Hamiltonian) dynamical models [9-11], which are also 
designed to capture the chaoticity underlying macroscopic 
irreversibility. The treatment of these models has led to some 
interesting developments, and Ref. [11] has provided a dynamically 
based picture of the hydrodynamical fluctuations about their 
nonequilibrium steady states. Moreover, in the case of a certain 
particular model, namely the symmetric exclusion process, it has been 
shown that the nonequilibrium steady state has long range density 
correlations [9-11] and that the probability distribution of its large 
scale density field is determined by an explicitly specified and highly 
nontrivial nonequilibrium generalization of its free energy [10, 11]. 
In the quantum regime, a natural dynamically based definition of 
nonequilibrium steady states of reservoir driven systems has been 
formulated [12, 13] at the microscopic level.
\vskip 0.2cm 
In the present article we set out a different approach to the 
subject, which is {\it quantum macrostatistical} in that is is centred 
on the hydrodynamical observables of reservoir driven quantum 
systems. This approach, which was briefly sketched in Ref. [14], 
parallels the one we have previously made to the nonequilibrium 
thermodynamics of conservative quantum systems [15, 16], where it 
yielded an extension of the Onsager reciprocity relations to a 
nonlinear regime. In general, the quantum macrostatistics is designed, 
like Onsager\rq s [17] irreversible thermodynamics and Landau\rq s 
fluctuating hydrodynamics [18], to form a bridge between the 
microscopic and macroscopic pictures of matter, rather than a deduction 
of the latter from the former. Indeed, accepting Boltzmann\rq s 
hypothesis of molecular chaos [19], we take the view that such a 
derivation is not even feasible for realistically interacting systems, 
since this chaos renders the microscopic equations of motion 
intractable over periods substantially longer than the intervals 
between successive collisions\footnote{$^{c}$}{This view is supported 
by the fact that the rigorous derivations of Boltzmann equations from 
the Hamiltonian dynamics of both classical [20] and quantum [21] 
systems are applicable only over microscopic times of the order of the 
interval between successive collisions of a particle. For longer times, 
the chaos bars the way to further analysis of the microscopic equations 
of motion.}. Thus, the microscopic equations of motion must necessarily 
be supplemented by further assumptions in order to interconnect the 
quantum and phenomenological properties of matter. In fact, the key 
physical assumptions of our macrostatistical project concern only very 
general, model-independent properties of many-particle systems. 
Specifically, they comprise 
\vskip 0.2cm\noindent
(A) an extension of Onsager\rq s regression hypothesis [17], to the 
effect that the hydrodynamical fluctuations about nonequilibrium steady 
states are governed by the same dynamical laws as the \lq small\rq\ 
perturbations of the hydrodynamical variables about their steady 
values;
\vskip 0.2cm\noindent
(B) a certain mesoscopic local equilibrium hypothesis; and
\vskip 0.2cm\noindent
(C) a chaoticity hypothesis for the nonconserved currents carried by 
the locally conserved hydrodynamical observables. 
\vskip 0.2cm\noindent
These assumptions may be regarded as the \lq axioms\rq\ of our theory. 
The physical considerations that underlie them will be discussed, along 
with their formulation, in Sections 4.1, 4.1 and 4.4. In fact, the 
hypothesis (C), like Boltzmann\rq s {\it Stosszahlansatz} and its 
subsequent developments [7-11], exploits the consequence of the very 
chaos that obstructs the analytical dynamics of realistically 
interacting many-particle systems.
\vskip 0.2cm
The principal results that we obtain by supplementing the Schroedinger 
dynamics of many-particle systems by the \lq axioms\rq\ (A)-(C), 
together with certain technical assumptions, are the following ones 
(I)-(III), which we claim to be new, at least on the level of a 
rigorous, general, model-independent quantum theory of nonequilibrium 
steady states.
\vskip 0.2cm\noindent 
(I) The spatial correlations of the hydrodynamical observables are 
generically of long range. This comprises a quantum mechanical 
generalization of that obtained from both rigorous microscopic 
treatments of certain classical stochastic models [9-11] and from 
heuristic treatments [23, 24] of Landau\rq s fluctuating hydrodynamics. 
Most importantly, it marks a qualitative 
difference between equilibrium and nonequilibrium steady states, 
since the hydrodynamical correlations in the former states are 
generically of short range, except at critical points.  
\vskip 0.2cm\noindent
(II) The transport coefficients satisfy a generalized, 
position-dependent version of the Onsager reciprocity relations. Thus, 
this result extends Onsager\rq s irreversible thermodynamics from the 
neighbourhood of equilibrium to that of nonequilibrium steady states. 
\vskip 0.2cm\noindent
(III) The hydrodynamical fluctuations execute a classical Gaussian 
Markov process, of a generalized Onsager-Machlup (OM) type [22]. Thus 
this result extends the OM theory from the regime of fluctuations about 
thermal equilibrium to that of fluctuations about nonequilibrium steady 
states. A similar result was obtained for certain classical stochastic 
models in Ref. [11].  
\vskip 0.2cm 
Let us now briefly describe the macrostatistical strategy we 
employ to obtain these results. We take our model to 
be an $N$-particle quantum system, ${\Sigma}$, that is confined to a 
bounded open connected region, ${\Omega}_{N}$, of a $d$-dimensional 
Euclidean space, $X$, and coupled at its boundary, 
${\partial}{\Omega}_{N}$, to an array, ${\cal R}$, of quantum 
mechanical reservoirs. ${\Sigma}$ is thus an open system, while the 
composite $({\Sigma}+{\cal R})$ is a conservative one. Since we shall 
have occasion to pass to thermodynamic and hydrodynamic limits where 
its particle number tends to infinity, we take $N$ to be a variable 
parameter of the system. We assume that its particle number density 
${\nu}:= N/{\rm Vol}({\Omega}_{N})$ is $N$-independent and that 
${\Omega}_{N}$ is the dilation by a factor $L_{N}$ of a fixed, $N$-
independent region ${\Omega}$ of unit volume. Thus 
${\Omega}_{N}=L_{N}{\Omega}:= 
{\lbrace}L_{N}x{\vert}x{\in}{\Omega}{\rbrace}$ and 
$$L_{N}=(N/{\nu})^{1/d}.\eqno(1.1)$$ 
For the hydrodynamic description of ${\Sigma}$, we take 
$L_{N}$ to be the unit of length. Thus, ${\Omega}$ is the region 
occupied by the system in the hydrodynamical picture.
\vskip 0.2cm
We assume that, in that picture, ${\Sigma}$ evolves according to a 
phenomenological law governing the evolution of a set of locally 
conserved classical fields 
$q_{t}(x)=\bigl(q_{1,t}(x),. \ .,q_{m,t}(x)\bigr)$, which 
correspond to the densities at position $x$ and time $t$ of the 
extensive thermodynamic variables\footnote{$^{d}$}{We provide a 
characterization of these variables in Section 2.2 along lines 
previously formulated in Ref. [15].} of the system. We denote the 
associated currents of $q_{t}(x)$ by $j_{t}(x)=\bigl(j_{1,t}(x),. \ 
.,j_{m,t}(x)\bigr)$. Thus, $q_{t}$ satisfies the local conservation law 
$${{\partial}q_{t}\over {\partial}t}+{\nabla}.j_{t}(x)=0.\eqno(1.2)$$ 
We assume that its phenomenological dynamics is governed by a 
constitutive equation of the form   
$$j_{t}(x)={\cal J}(q_{t};x),\eqno(1.3)$$ 
where ${\cal J}$ is a functional of the field $q_{t}$ and the 
position $x$. Thus, by Eqs. (1.2) and (1.3), $q_{t}$ evolves according 
to an autonomous law 
$${{\partial}q_{t}(x)\over {\partial}t}= 
{\cal F}(q_{t};x):=-{\nabla}.{\cal J}(q_{t};x),\eqno(1.4)$$ 
subject to boundary conditions determined by the reservoirs. 
We assume that this phenomenological law is invariant under scale 
transformations $x{\rightarrow}{\lambda}x, \ 
t{\rightarrow}{\lambda}^{k}t$ for some constant $k$. A simple 
example for which this assumption is valid, with $k=2$, is that 
of nonlinear diffusions, where ${\cal J}$ takes the form  
$${\cal J}(q_{t};x)= 
-{\tilde K}\bigl(q_{t}(x)\bigr){\nabla}q_{t}(x),\eqno(1.5)$$ 
${\tilde K}$ being an $m$-by-$m$ matrix $[{\tilde K}_{kl}]$, which 
acts by standard matrix multiplication on ${\nabla}q_{t}$. In 
this case, the phenomenological equation (1.4) takes the form 
$${{\partial}q_{t}\over {\partial}t}= 
{\nabla}.\bigl({\tilde K}(q_{t}){\nabla}q_{t}\bigr).\eqno(1.6)$$ 
We shall base some of our explicit calculations on this case and, 
in particular, we shall henceforth assume that the scaling 
exponent $k$ is equal to 2. A simple consequence of this assumption is 
that, since $L_{N}$ is the unit of length for the hydrodynamical 
picture, $L_{N}^{2}$ is the unit of time for this picture. 
\vskip 0.2cm 
We assume that, in general, the dynamics described by Eq. (1.4) 
is dissipative, in that the $m$-component field $q_{t}(x)$ 
relaxes eventually to a unique time-independent form $q(x)$, 
which thus corresponds to a steady hydrodynamical state. By Eq. 
(1.3), the corresponding steady $m$-component current, $j(x)$, 
is then ${\cal J}(q;x)$.  
\vskip 0.2cm 
By Eq. (1.4), the linearised equation of motion for \lq small\rq\ 
perturbations, ${\delta}q_{t}(x)$, of $q(x)$ is simply 
$${{\partial}\over {\partial}t}{\delta}q_{t}(x)= 
{\cal L}{\delta}q_{t}(x):= 
{{\partial}\over {\partial}{\lambda}} 
{\cal F}(q+{\lambda}{\delta}q_{t};x)_{{\vert}{\lambda}=0}, 
\eqno(1.7)$$ 
while, by Eq. (1.3), the corresponding increment in the $m$- 
component current $j(x)$ is  
$${\delta}j_{t}(x)={\cal K}{\delta}q_{t}(x):= 
{{\partial}\over {\partial}{\lambda}} 
{\cal J}(q+{\lambda}{\delta}q_{t};x)_{{\vert}{\lambda}=0}. 
\eqno(1.8)$$ 
We note that, by Eqs. (1.4), (1.7) and (1.8), 
$${\cal L}=-{\nabla}.{\cal K}.\eqno(1.9)$$ 
Further, in the case of nonlinear diffusions, it follows from the 
identification of the r.h.s.\rq s of Eqs. (1.4) and (1.6) that Eq. 
(1.7) yields the following formal equation for ${\cal L}$. 
$$[{\cal L}{\chi}](x)={\nabla}.\Bigl({\tilde K}\bigl(q(x)\bigr) 
{\nabla}{\chi}(x)+\bigl[{\tilde K}'\bigl(q(x)\bigr){\chi}(x)\bigr] 
{\nabla}q(x)\Bigr),\eqno(1.10)$$ 
where ${\chi}$ is a single column matrix function of position and 
${\tilde K}'(q)$ is the derivative of ${\tilde K}(q)$, i.e. its 
gradient with respect to $q$: thus  
$[{\tilde K}^{\prime}(q){\chi}(q)]_{kl}={\sum}_{r=1}^{m} 
[{\partial}{\tilde K}_{kl}(q)/{\partial}q_{r}]{\chi}_{r}(q)$.  
\vskip 0.2cm 
In order to relate the phenomenological dynamics given by Eqs. 
(1.4) and (1.7) to the underlying microscopic quantum mechanics of 
${\Sigma}$, we assume that $q_{t}(x)$ is the expectation value 
of a set of locally conserved quantum fields  
${\hat q}_{t}(x)=\bigl({\hat q}_{1,t}(x),. \ ., 
{\hat q}_{m,t}(x)\bigr)$ as rescaled for the hydrodynamical picture and 
in a limit in which $N$, and hence $L_{N}$, becomes infinite. 
Correspondingly, we formulate the fluctuations ${\xi}_{t}(x)$ of this 
$m$-component quantum field $q_{t}(x)$ 
about its mean on the same macroscopic scale and with a standard 
normalization, subject to the above-described assumptions (A)-(C). 
\vskip 0.2cm 
On this basis, we establish that ${\xi}_{t}$  
executes a Gaussian Markov process represented by a generalized 
Langevin equation of the form 
$${{\partial}\over {\partial}t}{\xi}_{t}(x)= 
{\cal L}{\xi}_{t}(x)+b_{t}(x),\eqno(1.11)$$ 
where $b_{t}(x)$ is a white noise whose 
autocorrelation function is of zero range with respect to position as 
well as time. Thus, ${\xi}_{t}$ executes a generalized Onsager-Machlup 
process. We employ this result to infer that the spatial 
correlations of the fluctuation field ${\xi}$ in nonequilibrium 
steady states are generically of long range. In this way we derive the 
above results (I)-(III) from our basic macrostatistical assumptions.   
\vskip 0.2cm  
We present our treatment as follows. In Section 2 we formulate
the quantum statistical thermodynamical model of the composite system 
$({\Sigma}+{\cal R})$ at both microscopic and macroscopic levels. This 
formulation provides general specifications of the nonequilibrium 
steady states of the model and also of the locally conserved quantum 
fields ${\hat q}_{t}$ and associated currents ${\hat j}_{t}$ pertinent 
to its hydrodynamic description. Here, in accordance with the general 
requirements of quantum field theory [25], we assume that these are 
distribution-valued operators. In Section 3 we relate 
the classical hydrodynamical variables, $q_{t}$ and $j_{t}$, and their 
fluctuations, ${\xi}_{t}$ and ${\eta}_{t}$, about a nonequilibrium 
steady state to these quantum fields and currents; and we obtain 
sufficient conditions for the fluctuations ${\xi}_{t}$ to execute a 
{\it classical} stochastic process. In Section 4 we formulate our 
regression and local equilibrium hypotheses for this process and note 
that these, together with the assumption of microscopic reversibility 
for the composite $({\Sigma}+{\cal R})$, yields a canonical extension 
of 
Onsager\rq s reciprocity relations to the nonlinear hydrodynamical 
regime. 
In Section 5 we extend our local equilibrium hypothesis to the 
fluctuating currents, ${\eta}_{t}$, and formulate our chaoticity 
hypothesis for these currents. We then establish that the assumptions 
of the regression hypothesis, local equilibrium and chaoticity imply 
the field ${\xi}_{t}$ executes a generalized Onsager-Machlup process 
represented formally by Eq. (1.11). In Section 6 we obtain an explicit 
formula for the two-point function for this process in terms of the 
equilibrium entropy density function and the transport coefficients of 
the system, and we infer therefrom that the static correlations of the 
hydrodynamical fluctuation field ${\xi}$ are generically of non-zero 
range on the macroscopic scale and hence of long (infinite!) range on 
the 
microscopic one. We conclude in Section 7 with some general 
observations about the results of this article and of their possible 
generalizations to less restrictive conditions than those assumed here. 
We leave the proofs of some technical Propositions to four Appendices.
\vskip 0.5cm
\centerline {\bf 2. The Quantum Model.}
\vskip 0.3cm
We take our model to be the open quantum system, ${\Sigma}$, briefly 
described 
in Section 1. Thus, ${\Sigma}$ is a system of $N$ particles, which 
occupies a 
bounded open connected region, ${\Omega}_{N}$, of a $d$-dimensional 
Euclidean 
space $X$ and is coupled at its surface, ${\partial}{\Omega}_{N}$, to 
an array, 
${\cal R}$, of reservoirs. Here ${\Omega}_{N}$ is the dilation by a 
factor 
$L_{N}$ of a region, ${\Omega}$, of unit volume and $L_{N}$ is given by 
Eq. (1), 
which represents the $N$-independence of the particle density of 
${\Sigma}$. We 
assume that the composite quantum system 
${\Sigma}^{(c)}:=({\Sigma}+{\cal R})$ 
is conservative and that all its interactions are invariant under 
spatial 
translations and rotations.
\vskip 0.3cm
{\bf 2.1. The Microscopic Picture.} We formulate this picture in 
standard 
operator algebraic terms, denoting the $C^{\star}$-algebras of bounded 
observables of ${\Sigma}$ and ${\Sigma}^{(c)}$ by ${\cal A}$ and ${\cal 
B}$ , 
respectively. We assume that ${\cal A}$ is a subalgebra of ${\cal B}$ 
and that 
it is isomorphic to the $W^{\star}$-algebra of bounded operators in a 
separable 
Hilbert space 
${\cal H}$, which comprises the square integrable functions $f(x_{1},. 
\
,x_{N};s_{1},. \ .,s_{N})$ (appropriately symmetrized or 
antisymmetrized) 
of the positions ${\lbrace}x_{j}{\rbrace}$ and the spins 
${\lbrace}s_{j} \ 
(={\pm}1){\rbrace}$ of its particles. The unbounded observables of 
${\Sigma}$ 
are represented by the unbounded self-adjoint operators affiliated
to ${\cal A}$, i.e. by those whose spectral projectors belong to
this algebra. The states of this system are represented by the density 
matrices 
in ${\cal H}$, and the expectation value of an observable, $A$, of 
${\Sigma}$ 
for the state ${\rho}$ is ${\rm Tr}({\rho}A)$. In general we denote 
this 
expectation value by ${\rho}(A){\equiv}{\langle}{\rho};A{\rangle}$, and 
we 
employ the corresponding notation for ${\Sigma}^{(c)}$.
\vskip 0.2cm
The Wigner time reversal operator, which serves
to reverse the velocities and spins of the particles of
${\Sigma}$, is defined to be the antilinear transformation of
${\cal H}$ given by the formula  
$$(Tf)(x_{1},. \ ,x_{N};s_{1},. \ .,s_{N})=
{\overline f}(x_{1},. \ ,x_{N};-s_{1},. \ .,-s_{N}) \
{\forall}f{\in}{\cal H},\eqno(2.1)$$
where the bar denotes complex conjugation. Thus, $T$ implements
an antiautomorphism ${\tau}_{\cal A}$ of ${\cal A}$, defined by
the formula
$${\tau}_{\cal A}A=TA^{\star}T \ {\forall} \ A{\in}{\cal A}.
\eqno(2.2)$$
\vskip 0.2cm
We assume that the dynamics of the composite system
${\Sigma}^{(c)}$ is given by a one-parameter
group, ${\lbrace}{\alpha}_{t}{\vert}t{\in}
{\bf R}{\rbrace}:={\alpha}({\bf R})$, of automorphisms of ${\cal B}$. 
Further, 
we assume that this dynamics is reversible, i.e. that ${\cal B}$ is
equipped with an antiautomorphism ${\tau}$, which reduces to
${\tau}_{\cal A}$ on ${\cal A}$ and implements time reversals
according to the prescription 
$${\tau}{\alpha}_{t}{\tau}={\alpha}_{-t}.\eqno(2.3)$$
The evolution of the observables of ${\Sigma}$ is given by the
isomorphisms of ${\cal A}$ into ${\cal B}$ obtained by the
restriction of ${\alpha}({\bf R})$ to the former algebra. 
\vskip 0.3cm
{\bf 2.2. Thermodynamic Variables and Potentials.} In order to
formulate the thermodynamic observables and potentials of
${\Sigma}$ we pass, for the moment, to the situation where it is
decoupled from the reservoirs ${\cal R}$ and thus becomes a
conservative system, whose dynamics is given by a one-parameter group, 
${\lbrace}{\alpha}_{t}^{(0)}{\vert}t{\in}{\bf R}{\rbrace}$, of 
automorphisms of 
${\cal A}$. In this situation, its canonical equilibrium
state, ${\rho}$, at inverse temperature ${\beta}$ is
characterized by the Kubo-Martin-Schwinger (KMS)
condition [26]
$${\langle}{\rho};[{\alpha}_{t}^{(0)}A_{1}]A_{2}{\rangle}=
{\langle}{\rho};A_{2}{\alpha}_{t+i{\hbar}{\beta}}^{(0)}A_{1}
{\rangle} \ {\forall} \ A_{1},A_{2} \ {\in} \ {\cal A}; \
t \ {\in} \ {\bf R}.\eqno(2.4)$$
Most importantly, this condition survives the thermodynamic limit
where $N$ tends to infinity and the particle density ${\nu}$
remains finite [26]. Moreover, in this limit\footnote{$^{e}$}{The model
of the infinite system is formulated, in a standard way, in terms
of its $C^{\star}$-algebra of quasi-local bounded observables
[15, 26-28]. Its states are then positive normalized linear functionals 
on that algebra.}, the system may support different states
that satisfy the condition. The set of these states is convex,
and its extremal elements may naturally be interpreted as the
pure equilibrium phases for the inverse temperature ${\beta}$
[15, 29].
\vskip 0.2cm
We assume that ${\Sigma}$ has a linearly independent set of
extensive conserved observables 
${\hat Q}=({\hat Q}_{1},. \ .,{\hat Q}_{n})$, which
intercommute\footnote{$^{f}$}{The assumption
of intercommutativity is not universally fulfilled. It is
violated, for example in the case where ${\hat Q}_{k}$ and ${\hat
Q}_{l}$, say, are different components of the magnetic moment of
${\Sigma}$. In such cases, some aspects of our treatment
would have to be refined.} up to surface effects and satisfy the
following condition of {\it thermodynamical completeness} [15]:-
{\it in the limit $N{\rightarrow}{\infty}$, the pure phases are
labelled by, i.e are in one-to-one correspondence with, the
expectation values $q_{1},. \ .,q_{m}$ of the global densities of 
${\hat Q}_{1},. \ .,{\hat Q}_{m}$, respectively.} The resultant set of 
classical, intensive thermodynamical variables of ${\Sigma}$ is then 
$q=(q_{1},. \ .,q_{m})$. In general, we take ${\hat Q}_{1}$ to be the 
Hamiltonian of the system: correspondingly, $q_{1}$ is its energy 
density.
\vskip 0.2cm
The equilibrium entropy density, in the limit
$N{\rightarrow}{\infty}$, is a function, $s$, of $q$, which
may be formulated by standard methods of quantum statistical
mechanics [15, 27]. The classical equilibrium thermodynamics
of the system is then governed by the form of $s(q)$. The
demand of thermodynamical stability ensures that this function
is concave. We define the thermodynamic conjugate of $q_{k}$ to
be ${\theta}_{k}={\partial}s(q)/{\partial}q_{k}$. Thus,
denoting the element $({\theta}_{1},. \ .,{\theta}_{m})$ of 
${\bf R}^{m}$ by ${\theta}$,
$${\theta}=s'(q),\eqno(2.5)$$
the derivative of $s(q)$, i.e. its gradient in $q$-space. 
Correspondingly, the 
second derivative, $s^{{\prime}{\prime}}(q)$, of this function
is the Hessian $[{\partial}^{2}s(q)/{\partial}q_{k}{\partial}q_{l}]$. 
We
assume throughout this treatment that the system is in a
single phase region, i.e. one where $s$ is infinitely
differentiable, where the function $q{\rightarrow}{\theta}(q)$
is invertible and where, for each value of $q$, the matrix
$s^{{\prime}{\prime}}(q)$ is invertible. We define
$$J(q):=-s^{{\prime}{\prime}}(q)^{-1},\eqno(2.6)$$
which, in view of the concavity of $s$, is a positive matrix.
\vskip 0.3cm
{\bf 2.3. The Reservoir System ${\cal R}$.} We assume that ${\cal R}$
comprises a set, ${\lbrace }{\cal R}_{J}{\rbrace}$, of spatially
disjoint reservoirs, such that each ${\cal R}_{J}$ is placed in contact
with a subregion ${\partial}{\Omega}_{N,J}$ of ${\partial}{\Omega}_{N}$ 
and
${\bigcup}_{J}{\partial}{\Omega}_{N,J}={\partial}{\Omega}_{N}$. 
Further, we 
assume that each ${\cal R}_{J}$ has a thermodynamically
complete set of global extensive conserved observables 
$({\hat Q}_{J,1},. \ .,{\hat Q}_{J,m})$ that are the natural
counterparts of ${\hat Q}_{1},. \ .,{\hat Q}_{m}$,
respectively, in that, when ${\Sigma}$ and 
${\cal R}_{J}$ are placed in contact, the observables
$({\hat Q}_{k}+{\hat Q}_{J,k})$ of ${\Sigma}^{(c)}$ are still 
conserved. 
Correspondingly, the
thermodynamic control variables of ${\cal R}_{J}$ conjugate to
$Q_{J}$ are the same as those of ${\Sigma}$, namely ${\theta}$. We
denote by ${\omega}_{J}({\theta}_{J})$ the equilibrium state of ${\cal 
R}_{J}$ 
for which its ${\theta}$-value is ${\theta}_{J}$. 
\vskip 0.3cm
{\bf 2.4. Nonequilibrium Steady States of ${\Sigma}^{(c)}$.}
Returning now to the situation where ${\Sigma}$ is an open
system, we assume that this is prepared according to the
following prescription. ${\Sigma}$ and the reservoirs 
${\lbrace}{\cal R}_{J}{\rbrace}$ are independently prepared in the 
remote past 
in states ${\rho}_{0}$ and 
${\lbrace}{\omega}_{J}({\theta}_{J}){\rbrace}$, 
respectively, where  ${\rho}_{0}$ is normal and the values of 
${\theta}_{J}$ 
generally varies from reservoir to reservoir: thus, in general, the 
reservoirs 
${\lbrace}{\cal R}_{J}{\rbrace}$ are not in equilibrium with one 
another. 
Following this preparation the systems ${\Sigma}$ and ${\cal R}$ are 
then 
coupled together and the resultant conservative composite evolves 
freely 
according to the dynamics governed by the automorphisms
${\alpha}({\bf R})$. We assume that, as established under suitable
asymptotically abelian conditions [12, 13], this dynamics acts
so as to drive the system\footnote{$^{g}$}{The same result has been 
also obtained constructively [30] for certain models, which however are 
too rudimentary for our present purposes. In particular, the version of 
${\Sigma}$ there is just an multi-level atom.} ${\Sigma}^{(c)}$ into a 
terminal
${\rho}_{0}$-independent state ${\phi} (=w^{\star}-{\rm 
lim}_{t\to\infty}
{\alpha}_{t}^{\star}[{\rho}_{0}{\otimes}_{J}{\omega}_{J}({\theta}_{J})] 
\ )$,
whose restriction to ${\cal A}$ is normal. This state is uniquely
determined by the states 
${\lbrace}{\omega}_{J}({\theta}_{J}){\rbrace}$.
Accordingly, we take ${\phi}$ to be the nonequilibrium steady
state of ${\Sigma}^{(c)}$ stemming from the
specified preparation, and we denote its GNS triple by 
$({\cal H}_{\phi},{\pi},{\Phi})$. 
\vskip 0.2cm
We note that, in view of the stationarity of
${\phi}$, the automorphisms ${\alpha}({\bf R})$ are implemented by a
unitary representation $U$ of ${\bf R}$ in ${\cal H}_{\phi}$ according
to the prescription [31]
$${\pi}({\alpha}_{t}B)=U_{t}{\pi}(B)U_{t}^{-1} \ {\forall} \
B{\in}{\cal B}, \ t{\in}{\bf R},\eqno(2.7)$$
where $U$ is defined by the formula 
$$U_{t}{\pi}(B){\Phi}={\pi}({\alpha}_{t}B){\Phi} \ {\forall} \
B{\in}{\cal B}, \ t{\in}{\bf R}.\eqno(2.8)$$
Since Eq. (2.8) is applicable to the subalgebra ${\cal A}$ of
${\cal B}$, the dynamics of the open system ${\Sigma}$, in the
normal folium of ${\phi}$, is given by the isomorphisms
implemented by $U$ of ${\pi}({\cal A})$ into ${\pi}({\cal B})$.
\vskip 0.2cm
Moreover, this prescription extends to the unbounded observables
of ${\Sigma}$ for the following reasons. Since the restriction
of ${\phi}$ to ${\cal A}$ is normal, so too, by Eq. (2.7), are
the representations ${\pi}$ and ${\pi}{\circ}{\alpha}_{t}$. It
follows [32] that these representations have canonical
extensions to the unbounded observables, $S$, of ${\Sigma}$
according to the prescription that, if
${\lbrace}E_{\lambda}{\rbrace}$ is the family of spectral
projectors of $S$, then those of ${\pi}(S)$ and
${\pi}({\alpha}_{t}S)$ are ${\lbrace}{\pi}(E_{\lambda}){\rbrace}$
and ${\lbrace}{\pi}({\alpha}_{t}E_{\lambda})=
U_{t}{\pi}(E_{\lambda})U_{t}^{-1}{\rbrace}$, respectively. Hence, the
extension of the formula (2.7) to the unbounded observables
takes the form
$${\pi}({\alpha}_{t}S)=U_{t}{\pi}(S)U_{-t}\eqno(2.9)$$
for all unbounded observables $S$ of ${\Sigma}$.
\vskip 0.3cm
{\bf 2.5. The Fields ${\hat q}$ and the Currents ${\hat j}$.} We assume 
that, in  
the GNS representation ${\pi}$ for the nonequilibrium steady state 
${\phi}$, the 
$m$-component extensive thermodynamical observable ${\hat Q}$ has a
position-dependent, locally conserved density 
${\hat q}(x)=\bigl({\hat q}_{1}(x),. \ .,{\hat q}_{m}(x)\bigr)$, with 
associated 
current density ${\hat j}(x)=({\hat j}_{1}(x),. \ .,{\hat j}_{m}(x))$. 
Thus the
${\hat q}_{k}$'s and ${\hat j}_{k}$'s are quantum fields and, in
accordance with the general requirements of quantum field theory
[25], we assume that they are distributions\footnote{$^{h}$}{In 
concrete cases, it is a simple matter to verify that the explicit 
formulae for these fields and currents are indeed distributions. For 
example, the number density operator at position $x$ is simply 
${\sum}_{r=1}^{N}{\delta}(x-x_{r})$, where $x_{r}$ is the position of 
the $r$\rq th particle.}, in the sense of L. Schwartz [33]. 
\vskip 0.2cm
We formulate these distributions in terms of the Schwartz spaces, 
${\cal D}({\Omega}_{N})$ and ${\cal D}_{V}({\Omega}_{N})$, of
real, infinitely differentiable scalar and ${\bf R}^{d}$-vector
valued functions, respectively, on $X$ with support in
${\Omega}_{N}$. We define ${\cal D}^{m}({\Omega}_{N})$ and 
${\cal D}^{m}_{V}({\Omega}_{N})$, respectively, to be the real
vector spaces given by their $m$'th topological powers, equipped
with the operations of binary addition and multiplication by real
numbers given by the formula
$${\lambda}(f_{1},. \ .,f_{m})+{\lambda}'(f_{1}',. \ ,f_{m}')=
({\lambda}f_{1}+{\lambda}'f_{1}',. \
.,{\lambda}f_{m}+{\lambda}'f_{m}')$$
$${\forall} \ {\lambda},{\lambda}'{\in}{\bf R}, \
f_{k},f_{k}'{\in}{\cal D}({\Omega}) \ {\rm or} \ {\cal 
D}_{V}({\Omega}), 
\ k=1,. \ .,m.$$
We denote by ${\cal D}^{{\prime}m}({\Omega}_{N})$ and 
${\cal D}_{V}^{{\prime}m}({\Omega}_{N})$ the topological dual vector
spaces of ${\cal D}^{m}({\Omega}_{N})$ and ${\cal 
D}_{V}^{m}({\Omega}_{N})$
respectively. Evidently, these are spaces of distributions (cf.
[33])). 
\vskip 0.2cm
We assume that the $m$-component fields ${\hat q}(x)$ and
${\hat j}(x)$ are operator valued elements of 
${\cal D}^{{\prime}m}({\Omega}_{N})$ and 
${\cal D}_{V}^{{\prime}m}({\Omega}_{N})$, respectively. For 
simplicitly, we also  
assume that the components, ${\hat q}_{k}$, of ${\hat q}$
are invariant under time-reversals\footnote{$^{i}$}{Standard
examples of time-reversal invariant ${\hat q}_{k}$'s are the
local number and energy densities of many-particle systems.}, i.e
that they commute with the Wigner time reversal operator $T$. 
\vskip 0.2cm
The algebraic properties of the field ${\hat q}(x)$ are governed
by the forms of the commutators $[{\hat q}_{k}(x),{\hat
q}_{l}(y)]$. We assume that these take the following form, which
is readily verified by the use of standard formulae in the case where
${\hat q}_{1}$ is the energy density of the system and the
remaining ${\hat q}_{k}$'s are the particle number densities for
the different species of its constituent particles.
$$[{\hat q}_{k}(x),{\hat q}_{l}(y)]=
i{\hbar}{\sum}_{r=1}^{m}c_{klr}
{\hat j}_{r}(x).{\nabla}{\delta}(x-y),\eqno(2.10)$$  
where the $c$'s are $N$-independent constants. This formula
evidently accords with our assumption that ${\hat Q}_{k}$'s
intercommute, up to surface effects: indeed it implies that their
commutators are just the integrals of currents over
${\partial}{\Omega}_{N}$.
\vskip 0.2cm 
We denote by ${\hat q}(f)$ and ${\hat j}(g)$ the \lq smeared fields'
obtained by integrating the distributions ${\hat q}$ and 
${\hat j}$ against test functions $f=(f_{1},. \ .,f_{m})$ and
$g=(g_{1},. \ .,g_{m})$, which belong to the spaces 
${\cal D}^{m}({\Omega}_{N})$ and ${\cal D}_{V}^{m}({\Omega}_{N})$
respectively. Thus
$${\hat q}(f)={\sum}_{k=1}^{m}\int_{{\Omega}_{N}}dx
{\hat q}_{k}(x)f_{k}(x)\eqno(2.11)$$
and
$${\hat j}(g)={\sum}_{k=1}^{m}\int_{{\Omega}_{N}}dx
{\hat j}_{k}(x).g_{k}(x).\eqno(2.12)$$
In general, these smeared fields are unbounded observables,
affiliated to the algebra ${\cal A}$. Therefore, by Eq. (2.7),
their evolutes at time $t$, which we denote by 
${\hat q}_{t}(f)$ and ${\hat j}_{t}(g)$, are their transforms
implemented by the unitary operator $U_{t}$. Thus, they are the
smeared fields corresponding to distribution valued operators
${\hat q}_{t}(x)=U_{t}{\hat q}(x)U_{t}^{-1}$ and 
${\hat j}_{t}(x)=U_{t}{\hat j}(x)U_{t}^{-1}$, respectively; and
the analogous statement may evidently be made for their
components ${\hat q}_{k,t}(x)$ and ${\hat j}_{k,t}(x)$. For
notational convenience, we shall sometimes
denote ${\hat q}_{t}(x), \ {\hat q}_{t}(f), \ {\hat j}_{t}(x)$
and ${\hat j}_{t}(g)$ by ${\hat q}(x,t), \ {\hat q}(f,t), \ 
{\hat j}(x,t)$ and ${\hat j}(g,t)$, respectively.
\vskip 0.2cm
We assume that the cyclic vector ${\Phi}$ for the state ${\phi}$
lies in the domain of all monomials in the smeared fields 
${\hat q}_{t}(f)$ and ${\hat j}_{t'}(g)$ and that the resultant
vector values of these monomials are continuous in the $f$'s,
$g$'s, $t$'s and $t'$'s.
\vskip 0.2cm
Since ${\hat j}$ is the current associated with ${\hat q}$, the
local conservation laws for the latter field may be expressed in
the form
$${\hat q}_{t}(f)-{\hat q}_{s}(f)=\int_{s}^{t}du
{\hat j}_{u}({\nabla}f) \ {\forall} \ t,s \ {\in} \ 
{\bf R}, \ f \ {\in} \ {\cal D}^{m}({\Omega}_{N}).\eqno(2.13)$$
\vskip 0.3cm
{\bf 2.6. The Hydrodynamical Scaling.} We assume that the
hydrodynamical observables of the open system
${\Sigma}$ comprise just the $m$-component field ${\hat q}$, as
viewed on the scale where the unit of length is $L_{N}$. Thus,
on this scale, the system is confined to the fixed region
${\Omega}$. Further, in accordance with our assumption, following
Eq. (1.6), that the macroscopic dynamics is invariant under
space-time scale transformations $x{\rightarrow}{\lambda}x, \
t{\rightarrow}{\lambda}^{2}t$, we assume that $L_{N}^{2}$ is the
unit of time corresponding to the length unit $L_{N}$. Hence, in
the normal folium of the nonequilibrium steady state ${\phi}$,
the $m$-component hydrodynamic field is represented by the
distribution valued operator 
$${\check q}_{t}(x):={\hat q}(L_{N}x,L_{N}^{2}t).
\eqno(2.14)$$ 
It follows from this equation and Eq. (2.11) that the smeared
hydrodynamic field obtained by integrating 
${\check q}_{t}(x)$ against a ${\cal D}^{m}({\Omega})$-class test
function $f$ is
$${\check q}_{t}(f)={\hat q}(f^{(N)},L_{N}^{2}t), \ {\forall}
\ f{\in}{\cal D}^{m}({\Omega}), \ t{\in}{\bf R},\eqno(2.15)$$ 
where $f^{(N)} \ ({\in}{\cal D}^{m}({\Omega}_{N}))$ is related
to $f$ according to the formula
$$f^{(N)}(x)=L_{N}^{-d}f(L_{N}^{-1}x) \ {\forall} \
x{\in}{\Omega}_{N}.\eqno(2.16)$$
\vskip 0.2cm
Since the scale transformation
$(x,t){\rightarrow}(L_{N}x,L_{N}^{2}t)$ sends ${\hat q}$ to
${\check q}$, it follows that the local conservation law (2.13), or 
formally
${{\partial}{\hat q}_{t}(x)/{\partial}t}=-{\nabla}.{\hat j}_{t}(x)$, 
will be 
preserved if its sends ${\hat j}_{t}(x)$ to ${\check j}_{t}(x)$, 
where
$${\check j}_{t}(x):=
L_{N}{\hat j}(L_{N}x,L_{N}^{2}t).\eqno(2.17)$$
It follows from this formula and Eq. (2.12) that the smeared field 
obtained by 
integrating ${\check j}_{t}(x)$ against a ${\cal D}_{V}^{m}({\Omega})$-
class 
test function $g$ is 
$${\check j}_{t}(g)={\hat j}(g^{(N)},L_{N}^{2}t),\eqno(2.18)$$
where
$$g^{(N)}(x)=L_{N}^{1-d}g(L_{N}^{-1}x).\eqno(2.19).$$ 
In view of Eqs. (2.15) and (2.18), it is a simple matter to confirm
that the local conservation law (2.13) retains its form in the
macroscopic description, i.e. that 
$${\check q}_{t}(f)-{\check q}_{s}(f)=\int_{s}^{t}du
{\check j}_{u}({\nabla}f) \ {\forall} \ t,s \ {\in} \ 
{\bf R}, \ f \ {\in} \ {\cal D}^{m}({\Omega}).\eqno(2.20)$$
\vskip 0.5cm
\centerline {\bf 3. Connection between the Quantum Picture, the 
Phenomenological Dynamics}
\vskip 0.2cm
\centerline {\bf and the Hydrodynamical Fluctuations}
\vskip 0.3cm
We now seek an inter-relationship between the quantum and
hydrodynamical properties of the macroscopic field ${\check
q}_{t}(x)$ and its current ${\check j}_{t}(x)$ in the limit where
$N$ tends to infinity. In order to formulate this limit, we shall
henceforth indicate the $N$-dependence of the quantum model by
attaching the superscript $(N)$ to the symbols ${\Sigma}, \ {\phi}, \ 
{\Phi}, \ U, \ {\hat q}, {\hat j}, \ {\check q}$ and ${\check j}$. The 
symbol ${\Sigma}$, without that superscript, will be reserved for the 
limiting case where $N$ becomes infinite. The symbol ${\Omega}$, on the 
other hand, will continue to represent the fixed region occupied by 
${\Sigma}^{(N)}$, in the 
{\it hydrodynamical} scaling, for all $N$.
\vskip 0.2cm
Our basic assumptions concerning the relationship between the
quantum and hydrodynamic pictures of the model are that, in the
limit $N{\rightarrow}{\infty}$, 
\vskip 0.2cm\noindent
(a) the stationary hydrodynamic fields $q(x)$ and $j(x)$ are the
expectation values of the quantum fields 
${\check q}_{t}^{(N)}(x)$ and ${\check j}_{t}^{(N)}(x)$,
respectively, for the steady state ${\phi}^{(N)}$; and 
\vskip 0.2cm\noindent
(b) the regressions of the fluctuations of these
fields are governed, in a sense that will be made precise in Section 4, 
by the same dynamical laws (1.7) and (1.8)
as the weak perturbations ${\delta}q_{t}(x)$ and
${\delta}j_{t}(x)$ of $q(x)$ and $j(x)$, respectively. 
\vskip 0.2cm\noindent
The regression hypothesis (b) is a natural generalization of that
proposed by Onsager [17] for fluctuations about equilibrium
states. We remark here that, since ${\cal D}'$ spaces are
complete, these assumptions imply that the classical fields
$q(x), \ j(x), \ {\delta}q_{t}(x)$ and ${\delta}j_{t}(x)$,
introduced in Section 1, are distributions.
\vskip 0.3cm
{\bf 3.1. Quantum Statistical Formulae the Hydrodynamical
Variables.} It follows immediately from our specifications that
the above assumption (a) signifies that
$$q(x)={\lim}_{N\to\infty}\bigl({\Phi}^{(N)},
{\check q}_{t}^{(N)}(x){\Phi}^{(N)}\bigr)\eqno(3.1)$$
and
$$j(x)={\lim}_{N\to\infty}\bigl({\Phi}^{(N)},
{\check j}_{t}^{(N)}(x){\Phi}^{(N)}\bigr),\eqno(3.2)$$
the $t$-independence of the r.h.s.'s of these formula being
guaranteed by the stationarity of ${\phi}^{(N)}$. 
\vskip 0.2cm
In order to bring the hydrodynamical description of the model
into line with thermodynamics, we introduce the field
${\theta}(x)=\bigl({\theta}_{1}(x),. \ ,{\theta}_{m}(x)\bigr)$,
conjugate to $q(x)$ as defined by the space-dependent version of
Eq. (2.5), namely
$${\theta}(x)=s'\bigl(q(x)\bigr).\eqno(3.3)$$ 
Since we are assuming that the system is perpetually in a single
phase region, and thus that the function $s'$ is invertible, it
follows from this formula that the fields $q(x)$ and
${\theta}(x)$ are in one-to-one correspondence.
\vskip 0.2cm
Turning now to the hydrodynamical equation (1.4), we see
immediately that the stationary field $q(x)$ is determined by the
requirement that ${\cal F}(q;x)=0$, together with the conditions
imposed by the ${\Sigma}^{(N)}-{\cal R}$ coupling at the boundary
${\partial}{\Omega}$ of ${\Omega}$. In order to specify these
conditions, we denote by ${\partial}{\Omega}_{J}$ the section of
${\partial}{\Omega}$ where ${\Sigma}^{(N)}$ is in contact with ${\cal
R}_{J}$. We then assume the following boundary condition.
\vskip 0.3cm
({\bf R}) On the section ${\partial}{\Omega}_{J}$ of the boundary of 
${\Sigma}$, the classical field ${\theta}(x)$ of this system takes the 
value
${\theta}_{J}$ of the control variables of the equilibrium state 
in which ${\cal R}_{J}$ is initially prepared. Thus the array of
reservoirs fixes the form of ${\theta}(x)$ and therefore of
$q(x)$ on ${\partial}{\Omega}$.
\vskip 0.3cm
This assumption signifies that, on the {\it hydrodynamic} time
scale and in the limit $N{\rightarrow}{\infty}$, the local
thermodynamical variables ${\theta}(x)$ of ${\Sigma}$
spontaneously take up the same values as the reservoir with which
this system is in contact at its boundary. The assumption is
fulfilled by the models of Refs. [9-11].
\vskip 0.3cm 
{\it Note on the Phenomenological Dynamics: ${\nabla}{\theta}$
as Driving Force.} In the general situation where the field $q_{t}$
is time-dependent, we define its thermodynamical conjugate to be
the field ${\theta}_{t}$ given by the space-time dependent
version of Eq. (2.5), namely
$${\theta}_{t}(x)=s'\bigl(q_{t}(x)\bigr).\eqno(3.4)$$
Thus, in view of our assumption that the system is perpetually
in a single phase region, the function $s'$ is invertible and the
phenomenological law (1.4) may be expressed in the form
$${{\partial}\over {\partial}t}q_{t}(x)=
{\nabla}.{\cal G}\bigl({\theta}_{t};x\bigr),\eqno(3.5)$$
where the functional ${\cal G}$ is determined by ${\cal J}$
according to the formula
$${\cal G}\bigl(s'(q_{t});x)=-{\cal J}(q_{t};x).\eqno(3.6)$$
In particular, in the case of nonlinear diffusion, it follows
from Eqs. (1.4), (1.5), (2.5) and (2.6) that this phenomenological law
reduces to the form
$${{\partial}\over {\partial}t}q_{t}(x)+
{\nabla}.\Bigl(K\bigl({\theta}_{t}(x)\bigr)
{\nabla}{\theta}_{t}(x)\Bigr)=0,\eqno(3.7)$$
where, in correspondence with the general relationship (2.5) between 
$q$ and ${\theta}$, 
$$K({\theta})={\tilde K}(q)J(q){\equiv}
{\tilde K}\bigl([s']^{-1}({\theta}\bigr)J
\bigl([s']^{-1}({\theta}\bigr).\eqno(3.8)$$
One sees immediately from Eq. (3.7) that the gradient of the
thermodynamical field ${\theta}_{t}$ acts as the hydrodynamical
driving force.
\vskip 0.3cm
{\bf 3.2. Linearized Perturbations of the Hydrodynamics.} In view
of our above remarks, ${\delta}q_{t}$ is a distribution that
satisfies Eq. (1.7) and vanishes on ${\partial}{\Omega}$. We
assume that the linear operator ${\cal L}$ appearing in that
equation is the generator of a one-parameter semigroup,
${\lbrace}T_{t}{\vert}t{\in}{\bf R}_{+}{\rbrace}:=T({\bf R}_{+})$, of 
transformations of
${\cal D}^{{\prime}m}({\Omega})$. The solution of Eq. (1.7) is
then
$${\delta}q_{t}=T_{t-s}{\delta}q_{s} \ {\forall} \
t{\geq}s{\geq}0.\eqno(3.9)$$
Correspondingly, by Eq. (1.8),
$${\delta}j_{t}={\cal K}{\delta}q_{t}=
{\cal K}T_{t-s}{\delta}q_{s} \ {\forall} \
t{\geq}s.\eqno(3.10)$$
Further, by Eq. (3.9) and the dissipativity condition stated in the 
paragraph before Eq. (1.7), 
$${\cal D}'-{\lim}_{t\to\infty}T_{t}{\psi}=0 \ {\forall} \ {\psi}
{\in}{\cal D}^{{\prime}m}({\Omega})\eqno(3.11)$$
or equivalently
$${\cal D}-{\lim}_{t\to\infty}T_{t}^{\star}f=0 \ {\forall} \ f
{\in}{\cal D}^{m}({\Omega}),\eqno(3.12)$$
where ${\lbrace}T_{t}^{\star}{\vert}t{\in}{\bf R}_{+}{\rbrace}$
is the one-parameter semigroup of transformations of ${\cal
D}^{m}({\Omega})$ dual to $T({\bf R}_{+})$. We denote its
generator by ${\cal L}^{\star}$, which is just the dual of 
${\cal L}$.
\vskip 0.3cm 
{\bf 3.3. The Hydrodynamical Fluctuation Fields.} We define the
quantum fields, 
${\xi}_{t}^{(N)}(x)=\bigl({\xi}_{1,t}^{(N)}(x),.. \ ,
{\xi}_{m,t}^{(N)}(x)\bigr)$ and
${\eta}_{t}^{(N)}=\bigl({\eta}_{1,t}^{(N)}(x),.. \ 
,{\eta}_{m,t}^{(N)}(x)\bigr)$, representing the fluctuations of
the hydrodynamically scaled field ${\check q}_{t}^{(N)}(x)$ and
the associated current ${\check j}_{t}^{(N)}(x)$, by the
formulae 
$${\xi}_{t}^{(N)}(x)=L_{N}^{d/2}
\bigl[{\check q}_{t}^{(N)}(x)-
\bigl({\Phi}^{(N)},
{\check q}_{t}^{(N)}(x){\Phi}^{(N)}\bigr)\bigr],\eqno(3.13)$$
and
$${\eta}_{t}^{(N)}(x)=L_{N}^{d/2}
\bigl[{\check j}_{t}^{(N)}(x)-
\bigl({\Phi}^{(N)},
{\check j}_{t}^{(N)}(x){\Phi}^{(N)}\bigr)\bigr],\eqno(3.14)$$
the normalization factor $L_{N}^{d/2}$ being natural for this
scaling. The corresponding smeared fields ${\xi}_{t}^{(N)}(f)$
and ${\eta}_{t}^{(N)}(g)$ are then the observables obtained by
integrating these fields against test functions $f \ 
({\in}{\cal D}^{m}({\Omega}))$ and $g \ ({\in}
{\cal D}_{V}^{m}({\Omega}))$, respectively. Thus, it follows from
Eqs. (2.20), (3.13) and (3.14) that ${\xi}_{t}^{(N)}$ satisfies the
local conservation law
$${\xi}_{t}^{(N)}(f)-{\xi}_{s}^{(N)}(f)=
\int_{s}^{t}du{\eta}_{u}^{(N)}({\nabla}f) \ {\forall} \
t,s \ {\in} \ {\bf R}, \ f \ {\in} \ {\cal D}^{m}({\Omega}_{N}).
\eqno(3.15)$$
\vskip 0.2cm
The dynamical properties of the fluctuation field
${\xi}_{t}^{(N)}$ are
encoded in the correlation functions
$$W^{(N)}(f^{(1)},. \ .,f^{(r)};t_{1},. \ .,t_{r})=
\bigl({\Phi}^{(N)},{\xi}_{t_{1}}^{(N)}(f^{(1)}). \ .
{\xi}_{t_{r}}^{(N)}(f^{(r)}){\Phi}^{(N)}\bigr).\eqno(3.16)$$ 
This formula, together with Eqs. (2.15) and (3.13), serves to express
$W^{(N)}$ in terms of the smeared fields ${\hat q}_{t}^{(N)}(f)$
of Section 2. Thus, in view of our stipulation there that the
actions on ${\Phi}^{(N)}$ of the monomials in these fields are
continuous in the $f$'s, and $t$'s, it follows that $W^{(N)}$ is
continuous in all its arguments. Further, it follows from the
stationarity of the state ${\phi}^{(N)}$ and the
self-adjointness of the observables ${\xi}_{t}^{(N)}(f)$ that
$$W^{(N)}(f^{(1)},. \ .,f^{(r)};t_{1}+a,. \ .,t_{r}+a)=
W^{(N)}(f^{(1)},. \ .,f^{(r)};t_{1},. \ .,t_{r}) \ {\forall} \
a{\in}{\bf R},\eqno(3.17)$$
and
$${\overline W^{(N)}(f^{(1)},. \ .,f^{(r)};t_{1},. \ .,t_{r})}=
W^{(N)}(f^{(r)},. \ .,f^{(1)};t_{r},. \ .,t_{1});\eqno(3.18)$$
while the positivity of ${\phi}^{(N)}$
implies that
$(A{\Phi}^{(N)},A{\Phi}^{(N)}){\geq}0$ for any polynomial $A$ in
the smeared fields ${\xi}_{t}^{(N)}(f)$. Thus choosing
$A={\sum}_{k=1}^{p}c_{k}{\xi}_{t_{k,1}}^{(N)}(f^{(k,1)}). \
.{\xi}_{t_{k,r_{k}}}^{(N)}(f^{(k,r_{k})})$, where the $c$'s are
complex constants and $p$ is finite,
$${\sum}_{k,l=1}^{p}{\overline c}_{k}c_{l}
W^{(N)}(f^{(k,r_{k})},. \ ,f^{(k,1)}f^{(l,1)},. \
,f^{(l,r_{l})};t_{k,r_{k}},. \ ,t_{k,1},t_{l,1},. \
,t_{l,r_{l}}){\geq}0.\eqno(3.19)$$  
\vskip 0.3cm
{\bf 3.4. Hydrodynamic Limit of the Fluctuation Process.} We
now assume that $W^{(N)}$ converges to a functional $W$ in the
hydrodynamic limit where $N{\rightarrow}{\infty}$, i.e.
that
$${\rm lim}_{N\to\infty}W^{(N)}(f^{(1)},. \ .,f^{(r)};t_{1},. \
.,t_{r})=W(f^{(1)},. \ .,f^{(r)};t_{1},. \ .,t_{r})$$
$${\forall} \ f^{(1)},. \ ,f^{(r)} \ {\in} \ 
{\cal D}^{m}({\Omega}),
 \ t_{1},. \ ,t_{r} \ {\in} \ {\bf R}, \ r{\in}{\bf N}.
\eqno(3.20)$$
Hence, in view of the continuity properties of $W^{(N)}$ and the
completeness of ${\cal D}'$ spaces, $W$ is continuous in the
$f$'s and measurable in the $t$'s. It is therefore a zero order 
distribution with respect to the latter variables [33]. Further, it 
follows
immediately from Eq. (3.20) that $W$ inherits the stationarity,
Hermiticity and positivity properties of $W^{(N)}$, as given by
Eqs. (3.17)-(3.19). Thus
$$W(f^{(1)},. \ .,f^{(r)};t_{1}+a,. \ .,t_{r}+a)=
W(f^{(1)},. \ .,f^{(r)};t_{1},. \ .,t_{r}) \ {\forall} \ a{\in}
{\bf R},\eqno(3.21)$$
$${\overline W(f^{(1)},. \ .,f^{(r)};t_{1},. \ .,t_{r})}=
W(f^{(r)},. \ .,f^{(1)};t_{r},. \ .,t_{1});\eqno(3.22)$$
and 
$${\sum}_{k,l=1}^{p}{\overline c}_{k}c_{l}
W(f^{(k,r_{k})},. \ ,f^{(k,1)},f^{(l,1)},. \
,f^{(l,r_{l})};t_{k,r_{k}},.
\ ,t_{k,1},t_{l,1},. \ ,t_{l,r_{l}}){\geq}0.\eqno(3.23)$$
\vskip 0.2cm
It follows from these properties that, by Wightman's
reconstruction theorem [25], $W$ corresponds precisely to the
quadruple $({\bf H},V,{\xi},{\Psi})$, where
\vskip 0.2cm\noindent
(a) ${\bf H}$ is a Hilbert space,
\vskip 0.2cm\noindent
(b) $V$ is a unitary representation of ${\bf R}$ in ${\bf H}$
such that $V_{t}$, the image of $t \ ({\in}{\bf R})$ under $V$,
is strongly measurable;
\vskip 0.2cm\noindent
(c) ${\xi}_{t}(x)$ is a Hermitian operator valued distribution,
of class ${\cal D}^{{\prime}m}({\Omega})$, in ${\bf H}$, which
implements the time translations of ${\xi}$, i.e.
$${\xi}_{t+s}(x)=V_{t}{\xi}_{s}(x)V_{t}^{-1};\eqno(3.24)$$
and
\vskip 0.2cm\noindent
(d) ${\Psi}$ is a vector in ${\bf H}$ that is invariant under
$V_{t}$ and cyclic with respect to the polynomials in the smeared
fields ${\xi}_{t}(f)$ obtained by integrating ${\xi}_{t}(x)$
against ${\cal D}^{m}({\Omega})$-class test functions $f$.
\vskip 0.2cm\noindent
The functional $W$ is then related to the smeared field
${\xi}_{t}(x)$ and the cyclic vector ${\Psi}$ by the formula
$$W(f^{(1)},. \ ,f^{(r)};t_{1},. \ ,t_{r})=
\bigl({\Psi},{\xi}_{t_{1}}(f^{(1)}). \
.{\xi}_{t_{r}}(f^{(r)}){\Psi}\bigr).\eqno(3.25)$$
\vskip 0.3cm
{\bf 3.5. Conditions for $W$ to represent a Classical Stochastic
Process.} The question of whether $W$ represents a classical stochastic 
process reduces to those of whether (a) it defines a quantum stochastic 
process in the sense of Ref. [34] and (b) this process has the abelian 
properties of a classical one. Now the condition (a) is fulfilled if 
the smeared Hermitian fields ${\xi}_{t}(f)$ are self-adjoint since, in 
this case, the unitary operators ${\lbrace}{\rm 
exp}\bigl(i{\xi}_{t}(f)\bigr)
{\vert}f{\in}{\cal D}^{m}({\Omega}){\rbrace}$ generate a
$W^{\star}$-algebra ${\cal N}_{t}$ and the correlation functions
$\bigl{\lbrace}({\Psi},F_{t_{1}}. \
.F_{t_{r}}{\Psi}){\vert}F_{t_{s}}{\in}
{\cal N}_{t_{s}}; \ s=1,. \ ,r\bigr{\rbrace}$
define a quantum stochastic process, as formulated in [34].
Further, the classicality condition\footnote{$^{j}$}{Here we consider
classical processes as special (abelian) cases of the quantum
ones.} (b) is simply that of the intercommutativity of the operators 
${\xi}_{t}(f)$.
\vskip 0.2cm 
The following proposition provides a sufficient condition for the
functional $W$ to represent a quantum stochastic process.
\vskip 0.3cm
{\bf Proposition 3.1.} {\it The functional $W$ uniquely defines
a quantum stochastic process ${\xi}$, indexed by 
${\cal D}^{m}({\Omega}){\times}{\bf R}$, if there is a bounded, 
positive
functional $(f,t){\rightarrow}F_{t}(f)$ on that product space 
such that} 
$${\vert}W(f^{(1)},. \ ,f^{(r)};t_{1},. \ ,t_{r}){\vert}{\leq}
r^{2}F_{t_{1}}(f^{(1)}). \ .F_{t_{r}}(f^{(r)}) \ {\forall} \ f^{(1)},.
\ .,f^{(r)} \ 
{\in}{\cal D}^{m}({\Omega}); \ t_{1},. \ .,t_{r}{\in}{\bf R}.
\eqno(3.26)$$
\vskip 0.3cm
{\bf Comment.} We shall subsequently establish in Prop. 6.1 that, under 
the assumptions of our scheme, the process ${\xi}$ is Gaussian. Since 
that result implies that the truncated $r$-point functions induced by 
$W$ all vanish and thus that Eq. (3.26) is satisfied, it signifies a 
consistency of our assumptions.
\vskip 0.3cm
{\bf Proof of Prop. 3.1.} As noted above, $W$ defines
a stochastic process if the Hermitian operators ${\xi}_{t}(f)$
are self-adjoint; and by Nelson's theorem [35], a sufficient
condition for this is that each of these fields has a dense
domain of analytic vectors. To prove that this is the case, subject to 
the assumption of Eq. (3.26), we note that it follows from that 
inequality and Eq. (3.25) that, for arbitrary $f, \ f^{(1)},. \ 
,f^{(r)}$ in 
${\cal D}^{m}({\Omega})$ and $t, \ t_{1},. \ ,t_{r}$ in ${\bf R}$,
$${\Vert}{\xi}_{t}(f)^{p}{\xi}_{t_{1}}(f^{(1)}).. \ 
.{\xi}_{t_{r}}(f^{(r)}){\Psi}{\Vert}{\leq}
(p+r)^{2}F_{t}(f)^{p}F_{t_{1}}(f^{(1)}).. \ .F_{t_{r}}(f^{(r)})$$
and therefore that the ${\bf H}$-valued function
$z({\in}{\bf C}){\rightarrow}{\sum}_{p=0}^{\infty}z^{p}
{\xi}_{t}(f)^{p}{\xi}_{t_{1}}(f^{(1)}). \
.{\xi}_{t_{r}}(f^{(r)}){\Psi}/p!$ has an infinite radius of
convergence. Hence, in view of the cyclicity of
${\Psi}$ with respect to the polynomials in the smeared fields 
${\lbrace}{\xi}_{t}(f){\rbrace}$, these fields are self-adjoint and 
therefore $W$ corresponds to a stochastic process. 
\vskip 0.3cm
We shall assume henceforth that $W$ does indeed define a stochastic 
process. In order to formulate a condition for its classicality, we 
introduce the following
definition.
\vskip 0.2cm
{\bf Definition 3.2.} (1) We define ${\cal P}$ (resp.
${\cal P}^{(N)}$) to be the set of polynomials in the smeared
fields ${\lbrace}{\xi}_{t}(f) \ \bigl({\rm resp.} \ 
{\xi}_{t}^{(N)}(f)\bigr){\vert}f{\in}{\cal D}^{m}({\Omega}),
\ t{\in}{\bf R}{\rbrace}$ and we define the bijection
$P{\rightarrow}P^{(N)}$ of ${\cal P}$ onto ${\cal P}^{(N)}$ by
the prescription that $P^{(N)}$ is the element of ${\cal P}^{(N)}$ 
obtained by 
replacing ${\xi}$ by ${\xi}^{(N)}$ in the formula for $P$.
\vskip 0.2cm
(2) For $P{\in}{\cal P}$ and $N{\in}{\bf N}$, we define the vector 
${\Psi}_{P}^{(N)} \ ({\in}{\cal H}_{{\phi}^{(N)}})$ by the formula 
$${\Psi}_{P}^{(N)}=P^{(N)}{\Phi}^{(N)}.\eqno(3.27)$$
\vskip 0.3cm
We now note that, by Eq. (3.25), the classicality condition that the 
operators ${\xi}_{t}(f)$ intercommute is equivalent to the invariance 
of 
$W(f^{(1)},. \ ,f^{(k)};t_{1},. \ ,t_{n})$ under the permutations
$$(f^{(r)},t_{r}){\rightleftharpoons}(f^{(r+1)},t_{r+1});$$ 
and by Def. (3.2) and Eqs. (3.12), (3.16), (3.20), this latter 
condition may be expressed in the  form
$${\lim}_{N\to\infty}\bigl({\Psi}_{P}^{(N)},
[{\xi}_{t}^{(N)}(f),{\xi}_{t'}^{(N)}(f')]
{\Psi}_{P}^{(N)}\bigr)=0$$
$${\forall} \ P{\in}{\cal P},  \  
f,f'{\in}{\cal D}^{m}({\Omega}), \ t,t'{\in}{\bf R}.$$
Moreover, we can set $t'=0$ here without loss of generality, since 
${\Phi}^{(N)}$ is invariant under $U_{t}^{(N)}$ and therefore, by Eq. 
(2.14), 
Def. 3.2 and the definition of ${\xi}_{t}^{(N)}$, the manifold 
${\cal P}^{(N)}{\Phi}^{(N)}$ is stable under this unitary 
transformation. Consequently, we have the following proposition, whose 
significance we shall discuss below. 
\vskip 0.3cm
{\bf Proposition 3.3.} {\it Under the above assumptions, the
process ${\xi}$ is classical if and only if ${\xi}_{t}^{(N)}(f)$ 
satisfies the 
condition that
$${\rm lim}_{N\to\infty}\bigl({\Psi}_{P}^{(N)},[{\xi}_{t}^{(N)}(f),
{\xi}^{(N)}(f^{\prime})]{\Psi}_{P}^{(N)}\bigr)=0 \ {\forall} 
P{\in}{\cal P}, \ f,f^{\prime}{\in}{\cal D}^{m}({\Omega}), \  
t{\in}{\bf R}.\eqno(3.28)$$}
\vskip 0.3cm
{\bf Comment.} In order to relate the condition (3.28) to the 
microscopic 
picture, we infer from Eqs. (2.10), (2.14)-(2.19) and (3.13) that this 
condition signifies the following. 
\vskip 0.2cm\noindent
(1) In the case where $t{\neq}0$,
$${\rm lim}_{N\to\infty}L_{N}^{d}
{\sum}_{k,l=1}^{m}\int_{{\Omega}^{2}}dxdy\bigl({\Psi}_{P}^{(N)},
[{\hat q}_{k}(L_{N}x,L_{N}^{2}t),{\hat 
q}_{l}(L_{N}y)]{\Psi}_{P}^{(N)}\bigr)
f_{k}(x)f_{l}^{\prime}(y)=0$$
$${\forall} \ f,f^{\prime}{\in}{\cal D}^{m}({\Omega}), \ P{\in}{\cal 
P},\eqno(3.29)$$
which is evidently a space-time asymptotic abelian condition on the 
field 
${\hat q}$.  
\vskip 0.2cm\noindent
(2) In the case where $t=0$, 
$${\rm lim}_{N\to\infty}L_{N}^{-2}\bigl({\Psi}_{P}^{(N)},
{\check j}^{(N)}(g_{f,f^{\prime}}){\Psi}_{P}^{(N)}\bigr)=0$$
$${\forall} \ f,f^{\prime}{\in}{\cal D}^{m}({\Omega}), \ P{\in}{\cal 
P},
\eqno(3.30)$$
where
$g_{f,f^{\prime}}$ is the element of ${\cal D}_{V}^{m}({\Omega})$ whose 
$r$'th component is 
$$g_{f,f^{\prime};r}=
{\sum}_{kl}{\hbar}c_{rkl}f_{k}{\nabla}f_{l}^{\prime}.\eqno(3.31)$$
Thus, Eq. (3.30) signifies the avoidance of the catastrophe whereby, 
for fixed $P{\in}{\cal P}$, the expectation value of the smeared 
hydrodynamically scaled current ${\check j}^{(N)}(g_{f,f^{\prime}})$ in 
the vector state ${\Psi}_{P}^{(N)}$ would grow as rapidly as 
$L_{N}^{2}$ with increasing $N$. 
\vskip 0.5cm
\centerline {\bf 4. The Stochastic Process ${\xi}$: Regression
and Local Equilibrium Hypotheses}
\vskip 0.2cm
\centerline {\bf and the Generalized Onsager Relations}
\vskip 0.3cm
We now assume that the conditions of Props. 3.1 and 3.3 are
fulfilled and hence that ${\xi}$ is a classical stochastic
process, indexed by ${\bf R}{\times}{\cal D}^{m}({\Omega})$. In
a standard way, we denote the expectation functional of the
random variables for this process by $E$. Thus, by Eq. (3.25),
$$E\bigl({\xi}_{t_{1}}(f^{(1)}). \ .{\xi}_{t_{r}}(f^{(r)})\bigr)=
\bigl({\Psi},{\xi}_{t_{1}}(f^{(1)}). \
.{\xi}_{t_{r}}(f^{(r)}){\Psi}\bigr) \ {\forall} \ t_{1},. \
.,t_{r} \ {\in} \ {\bf R}, \ f^{(1)},. \ .,f^{(r)} \ {\in} \ 
{\cal D}^{m}({\Omega}).\eqno(4.1)$$
We note that, by Eqs. (3.20), (3.25) and (4.1), the process
${\xi}^{(N)}$ converges to ${\xi}$, i.e. its correlation
functions converge to the corresponding ones for ${\xi}$,
as $N{\rightarrow}{\infty}$. Further, in view of the
observation following Eq. (3.20), the correlation function
$E\bigl({\xi}_{t_{1}}(f^{(1)}). \ .{\xi}_{t_{r}}(f^{(r)})\bigr)$
is continuous with respect to the $f$'s and measurable with respect to 
the $t$'s.
\vskip 0.3cm
{\it Conditional Expectations.} For any random variable $F$ of
the ${\xi}$-process and for $t{\in}{\bf R}$, we denote the
conditional expectations of $F$ with respect to the
${\sigma}$-algebras generated by
${\lbrace}{\xi}_{t}(f){\vert}f{\in}
{\cal D}^{m}({\Omega})){\rbrace}$ and 
${\lbrace}{\xi}_{t'}(f){\vert}t'{\leq}t, \ 
f{\in}{\cal D}^{m}({\Omega}){\rbrace}$ by $E(F{\vert}{\xi}_{t})$
and
$E(F{\vert}{\xi}_{{\leq}t})$, respectively. 
\vskip 0.3cm
{\bf 4.1. The Regression Hypothesis.} This hypothesis is just the
canonical generalization of that assumed by Onsager [17] for
fluctuations about equilibrium states. Its essential import is 
that the evolution of a small hydrodynamical deviation from a steady 
state does not depend on whether the deviation has arisen from a 
spontaneous fluctuation or from a weak perturbation of the 
system\footnote{$^{k}$}{As in Onsager\rq s theory, the assumption of 
this equivalence between the consequences of fluctuations and weak 
perturbations is not quite innocuous, since the modifications of the 
variables $q$ due to the former are $O(N^{-1/2})$, whereas those due to 
the latter are of order of a different small parameter that represents 
the strength of the perturbation.}. Thus, in mathematical terms, the 
regression hypothesis asserts that, for fixed $s$ and $t{\geq}s$, the 
evolution of $E({\xi}_{t}{\vert}{\xi}_{s})$ is governed by the
same law as that of the linearised perturbation ${\delta}q_{t}$
of the deterministic trajectory $q_{t}$, i.e., by Eq. (3.9), that
$$E\bigl({\xi}_{t}(f){\vert}{\xi}_{s}\bigr)=
[T_{t-s}{\xi}_{s}](f){\equiv}{\xi}_{s}(T_{t-s}^{\star}f) \
{\forall} \ t{\geq}s.\eqno(4.2)$$
Hence, since Nelson's forward time derivative [36] of ${\xi}_{t}(f)$ is
defined to be
$$D{\xi}_{t}(f):={\lim}_{u{\rightarrow}+0}u^{-1}
E\bigl({\xi}_{t+u}(f)-
{\xi}_{t}(f){\vert}{\xi}_{t}\bigr)\eqno(4.3)$$
and, since ${\cal L}$ is the generator of $T({\bf R}_{+})$, it follows 
that
$$D{\xi}_{t}(f)={\cal L}{\xi}_{t}(f).\eqno(4.4)$$
Further, defining the static two-point function 
$W_{S}:{\cal D}^{m}({\Omega}){\times}
{\cal D}^{m}({\Omega}){\rightarrow}{\bf R}$ by the formula
$$W_{S}(f,f')=E\bigl({\xi}(f){\xi}(f')\bigr) \ {\forall} \ 
f,f^{\prime}{\in}{\cal D}^{m}({\Omega}),\eqno(4.5)$$
it follows from Eq. (4.2) and the stationarity of the ${\xi}$-
process that
$$E\bigl({\xi}_{t}(f){\xi}_{t'}(f')\bigr)=
W_{S}(T_{t-t'}^{\star}f,f') \ {\forall} \ f,f^{\prime}{\in}{\cal 
D}^{m}({\Omega}), \ t,t^{\prime}({\leq}t){\in}{\bf R}.\eqno(4.6)$$
\vskip 0.3cm
{\bf 4.2. Local Equilibrium Conditions.} Our next assumption asserts 
essentially that, in a nonequilibrium steady state, the statistical 
properties of the fluctuation field ${\xi}$ in a \lq small\rq\ 
neighbourhood, ${\cal N}(x)$, of an arbitrary point $x \ 
({\in}{\Omega})$ simulate those enjoyed by these fields in the true 
equilibrium state corresponding to the value $q(x)$ of the 
thermodynamic variable $q$. This is a mesoscopic local equilibrium 
condition, since it involves only the fluctuation field ${\xi}$ and is 
thus weaker than that of microscopic local equilibrium [37], which 
would signify that the microstate of ${\Sigma}$ in ${\cal N}(x)$ 
simulated the equilibrium microstate corresponding to $q(x)$ there. 
Here we note that even this stronger condition has been shown to be 
fulfilled [38] by systems of fermions for which an Eulerian 
hydrodynamics has been established. Moreover, it may be expected to 
ensue more generally from the fact that the ratio of the hydrodynamic 
time scale to that of the microscopic processes (collisions etc.) is 
infinite, since that implies that local values of the hydro-
thermodynamic variables $q$ change negligibly in the time taken for the 
latter processes to generate equilibrium in macroscopically small 
spatial regions.     
\vskip 0.2cm
In order to precisely specify our mesoscopic local equilibrium 
hypothesis, we start by formulating the relevant properties of 
hydrodynamical fluctuations about true equilibrium states for which the 
stationary classical field $q(x)$ is assumed to be uniform.
\vskip 0.3cm
{\it Equilibrium Fluctuations.} We recall that, for a {\it
finite} system, the equilibrium probability distribution
function, $P$, for macroscopic observables $A$ is determined by the
entropy $S(A)$ according to the Einstein formula 
$$P(A)={\rm const.}{\rm exp}\bigl(S(A)\bigr),$$ 
and this serves to relate the static correlation functions for
the fluctuations of these observables to the thermodynamics of
the system. The generalization of this relation to infinite
systems has been derived by a quantum statistical treatment [15,
Ch.7, Appendix C] of equilibrium states and takes the form
$$E_{eq}\bigl({\xi}(f){\xi}(f^{\prime})\bigr)=
\bigl(f,J(q)f^{\prime}\bigr), \
{\forall} \ f,f^{\prime} \ {\in} \ 
{\cal D}^{m}({\Omega}),\eqno(4.7)$$
where $E_{eq}$ is the equilibrium expectation functional for
the fluctuation process, $J(q)$ is defined by Eq. (2.6) and
$(.,.)$ is the inner product on ${\cal D}^{m}({\Omega})$ defined
by the formula
$$(f,f^{\prime})={\sum}_{k=1}^{m}\int_{\Omega}dx
f_{k}(x)f_{k}^{\prime}(x).\eqno(4.8)$$
It follows from Eqs. (4.2) and (4.7) that
$$E_{eq}\bigl({\xi}_{t}(f){\xi}_{s}(f^{\prime})\bigr)=
\bigl(T_{t-s}^{\star}f,J(q)f^{\prime}\bigr) \ {\forall} \ 
f,f^{\prime} \ {\in} \ {\cal D}^{m}({\Omega}), \
t{\geq}s.\eqno(4.9)$$
Further, recalling the assumption, introduced in
Section 2.5, of the invariance of the quantum field
${\hat q}^{(N)}(x)$ under the time-reversal antiautomorphism
${\tau}$ and assuming that the equilibrium state\footnote{$^{l}$}{The 
same 
assumption would not be valid for nonequilibrium states, since these 
generally 
carry currents of odd parity with respect to time reversals.} 
${\phi}_{eq}^{(N)}$ of $({\Sigma}^{(N)}+{\cal R})$ is likewise
time-reversal invariant, it follows from the stationarity of this
state and Eq. (3.13) that
$${\langle}{\phi}_{eq}^{(N)};
{\xi}_{t}^{(N)}(f){\xi}^{(N)}(f^{\prime}){\rangle}= 
{\langle}{\phi}_{eq}^{(N)};
{\xi}^{(N)}(f^{\prime}){\xi}_{-t}^{(N)}(f){\rangle}=
{\langle}{\phi}_{eq}^{(N)};
{\xi}_{t}^{(N)}(f^{\prime}){\xi}^{(N)}(f){\rangle}.$$
On passing to the limit of this equation as
$N{\rightarrow}{\infty}$, we see that
$$E_{eq}\bigl({\xi}_{t}(f){\xi}(f^{\prime})\bigr)=
E_{eq}\bigl({\xi}_{t}(f^{\prime}){\xi}(f)\bigr);$$
and therefore, by Eq. (4.9), that
$$E_{eq}\bigl({\xi}(T_{t}^{\star}f){\xi}(f^{\prime})\bigr)=
E_{eq}\bigl({\xi}(T_{t}^{\star}f^{\prime}){\xi}(f\bigr),
\ {\forall} \ t{\geq}0.$$
Consequently, since ${\cal L}^{\star}$ is the generator of
$T^{\star}({\bf R}_{+})$,
$$E_{eq}\bigl({\xi}({\cal L}^{\star}f){\xi}(f^{\prime})\bigr)=
E_{eq}\bigl({\xi}({\cal L}^{\star}f^{\prime}){\xi}(f)\bigr) \ {\forall} 
\ 
f,f^{\prime}{\in}{\cal D}^{m}({\Omega}).\eqno(4.10)$$
\vskip 0.3cm
{\it Local Form of Equilibrium Correlations.} We formulate
the local properties of the equilibrium fluctuations in terms of
test functions that are highly localised around an arbitrary
point $x_{0}$ of ${\Omega}$. Specifically, for $f{\in}
{\cal D}^{m}({\Omega}), \ x_{0}{\in}{\Omega}$ and
${\epsilon}{\in}{\bf R}_{+}$, we define the function
$f_{x_{0},{\epsilon}}$ on the Euclidean space $X$ by the formula
$$f_{x_{0},{\epsilon}}(x)=
{\epsilon}^{-d/2}f\bigl({\epsilon}^{-1}(x-x_{0})\bigr) \
{\forall} \ x_{0}{\in}{\Omega}, \ f{\in}
{\cal D}^{m}({\Omega}).\eqno(4.11)$$
Since ${\Omega}$ is a bounded open subregion of $X$, it follows
that the restriction of $f_{x_{0},{\epsilon}}$ to ${\Omega}$
belongs to the space ${\cal D}^{m}({\Omega})$ for sufficiently
small ${\epsilon}$. In this case, we may take Eq. (4.11) to
define a transformation
$f{\rightarrow}f_{x_{0},{\epsilon}}$ of ${\cal D}^{m}({\Omega})$,
with ${\epsilon}$ representing the degree of localization of the
latter function about the point $x_{0}$.
\vskip 0.2cm
We now note that, by Eqs (4.8) and (4.11), the r.h.s. of Eq.
(4.7) is invariant under the
transformation $f{\rightarrow}f_{x_{0},{\epsilon}}$ and
therefore it follows from that equation that the equilibrium
fluctuations enjoy the {\it local} property given by the formula 
$${\rm lim}_{{\epsilon}{\downarrow}0}
E_{eq}\bigl({\xi}(f_{x_{0},{\epsilon}})
{\xi}(f_{x_{0},{\epsilon}}^{\prime})\bigr)=(f,J(q)f^{\prime}) \
{\forall} \ x_{0}{\in}{\Omega}, \ f,f^{\prime}{\in}{\cal 
D}^{m}({\Omega}).
\eqno(4.12)$$
Further, in the case of nonlinear diffusion, it follows from Eq. (1.10) 
that, for perturbations of the equilibrium state, 
${\cal L}={\tilde K}(q){\Delta}$, with $q$ constant. Hence, for 
fluctuations about equilibrium, it follows from Eq. (4.7) that both 
sides of Eq. (4.10) are 
invariant under the transformation 
$f{\rightarrow}f_{x_{0},{\epsilon}}, \ 
f^{\prime}{\rightarrow}f_{x_{0},{\epsilon}}^{\prime}, \ 
E_{eq}{\rightarrow}{\epsilon}^{2}E_{eq}$, and consequently  
$${\rm lim}_{{\epsilon}{\downarrow}0}{\epsilon}^{2}
E_{eq}\bigl({\xi}({\cal L}^{\star}f_{x_{0},{\epsilon}})
{\xi}(f_{x_{0},{\epsilon}}^{\prime})\bigr)=
{\rm lim}_{{\epsilon}{\downarrow}0}{\epsilon}^{2}
E_{eq}\bigl({\xi}({\cal L}^{\star}f_{x_{0},{\epsilon}}^{\prime})
{\xi}(f_{x_{0},{\epsilon}}\bigr) \
{\forall}x_{0}{\in}{\Omega}.\eqno(4.13)$$
\vskip 0.3cm
{\it Local Equilibrium Conditions for Nonequilibrium Steady States.} We
now assume that, for these states, the natural counterparts of the 
local conditions (4.12) and (4.13) still hold, i.e. that
$${\rm lim}_{{\epsilon}{\downarrow}0}
E\bigl({\xi}(f_{x_{0},{\epsilon}})
{\xi}(f_{x_{0},{\epsilon}}^{\prime})\bigr)=
\bigl(f,J(q(x_{0}))f^{\prime}\bigr)
\ {\forall} \ x_{0}{\in}{\Omega}, \ f,f^{\prime}{\in}{\cal 
D}^{m}({\Omega})
\eqno(4.14)$$
and
$${\rm lim}_{{\epsilon}{\downarrow}0}{\epsilon}^{2}
E\bigl({\xi}({\cal L}^{\star}f_{x_{0},{\epsilon}})
{\xi}(f_{x_{0},{\epsilon}}^{\prime})\bigr)=
{\rm lim}_{{\epsilon}{\downarrow}0}{\epsilon}^{2}
E\bigl({\xi}({\cal L}^{\star}f_{x_{0},{\epsilon}}^{\prime})
{\xi}(f_{x_{0},{\epsilon}})\bigr) \
{\forall} \ x_{0}{\in}{\Omega}, \ f,f^{\prime}{\in}{\cal 
D}^{m}({\Omega}).
\eqno(4.15)$$ 
These are our local equilibrium conditions, which manifestly concern 
the fluctuation field ${\xi}$ only.
\vskip 0.3cm
{\bf 4.3. Generalized Onsager Reciprocity Relations.} The following 
proposition 
represents a generalization of the Onsager reciprocity relations to 
nonequilibrium steady states of the nonlinear diffusion process.
\vskip 0.3cm
{\bf Proposition 4.1.} {\it Under the assumption of the regression and 
local 
equilibrium hypotheses, the transport coefficients of the nonlinear 
diffusion 
process satisfy the position-dependent Onsager relations}
$$K_{kl}\bigl({\theta}(x)\bigr)=K_{lk}\bigl({\theta}(x)\bigr) \ 
{\forall} \ x{\in}{\Omega}, \ k,l{\in}[1,m].\eqno(4.16)$$
\vskip 0.3cm
{\bf Proof.} Since we employ the same argument as that for 
nonequilibrium states of conservative systems in Ref. [15, Ch. 7], we 
shall just sketch the proof here. We start by introducing the linear 
transformation ${\cal L}_{0}$ of 
${\cal D}^{m}({\Omega})$ by the formula
$${\cal L}_{0}:={\tilde K}\bigl(q(x_{0})\bigr){\Delta}.\eqno(4.17)$$
It then follows, after some manipulation, from Eqs. (1.10), (3.8), 
(4.14) and (4.17), together with the continuity properties of the 
functions ${\tilde K}, \ J$ and $q$, that
$${\rm lim}_{{\epsilon}{\downarrow}0}{\epsilon}^{2}
E\bigl({\xi}([{\cal L}^{\star}-{\cal L}_{0}^{\star}]
f_{x_{0},{\epsilon}}){\xi}(f_{x_{0},{\epsilon}}^{\prime})\bigr)=0 \ 
{\forall} \ f,f^{\prime}{\in}{\cal D}^{m}{\Omega}), \ 
x_{0}{\in}{\Omega}.\eqno(4.18)$$
This implies that ${\cal L}$ may be replaced by ${\cal L}_{0}$ in Eq. 
(4.15), i.e. that 
$${\rm lim}_{{\epsilon}{\downarrow}0}{\epsilon}^{2}
E\bigl({\xi}({\cal L}_{0}^{\star}f_{x_{0},{\epsilon}})
{\xi}(f_{x_{0},{\epsilon}}^{\prime})\bigr)=
{\rm lim}_{{\epsilon}{\downarrow}0}{\epsilon}^{2}
E\bigl({\xi}({\cal L}_{0}^{\star}f_{x_{0},{\epsilon}}^{\prime})
{\xi}(f_{x_{0},{\epsilon}})\bigr) \
{\forall} \ f,f^{\prime}{\in}{\cal D}^{m}({\Omega}), \ 
x_{0}{\in}{\Omega}.\eqno(4.19)$$
Further, since, by Eqs. (4.11) and (4.17),
$${\epsilon}^{2}{\cal L}_{0}f_{x_{0},{\epsilon}}=
[{\cal L}_{0}f]_{x_{0},{\epsilon}},$$
Eq. (4.19) reduces to the form
$${\rm lim}_{{\epsilon}{\downarrow}0}
E\bigl({\xi}([{\cal L}_{0}f]_{x_{0},{\epsilon}})
{\xi}(f_{x_{0},{\epsilon}}^{\prime})\bigr)=
{\rm lim}_{{\epsilon}{\downarrow}0}
E\bigl({\xi}([{\cal L}_{0}f^{\prime}]_{x_{0},{\epsilon}})
{\xi}(f_{x_{0},{\epsilon}})\bigr) \
{\forall} \ f,f^{\prime}{\in}{\cal D}^{m}({\Omega}), \ 
x_{0}{\in}{\Omega}.$$
It follows from this equation, together with Eqs. (3.8), (4.14) and 
(4.17) that
$$\Bigl({\Delta}f,K\bigl({\theta}(x_{0})\bigr)f^{\prime}\Bigr)=
\Bigl({\Delta}f^{\prime},K\bigl({\theta}(x_{0})\bigr)f\Bigr), \ 
{\forall} \ f,f^{\prime}{\in}{\cal D}^{m}({\Omega}), \ 
x_{0}{\in}{\Omega}. \eqno(4.20)$$
Further, since, by Eq. (4.8), 
$$({\Delta}f,f^{\prime}){\equiv}({\Delta}f^{\prime},f) \ {\forall} \ 
f,f'{\in}
{\cal D}^{m}({\Omega}),$$
and since the actions of ${\Delta}$ and $K\bigl({\theta}(x_{0})\bigr)$ 
on 
${\cal D}^{m}({\Omega})$ intercommute, Eq. (4.20) is equivalent to the 
following formula. 
$$\Bigl({\Delta}f,K\bigl({\theta}(x_{0})\bigr)f^{\prime}\Bigr)=
\Bigl({\Delta}f,K^{\star}\bigl({\theta}(x_{0})\bigr)f^{\prime}\Bigr) \ 
{\forall} \ f,f^{\prime}{\in}{\cal D}^{m}({\Omega}), \ 
x_{0}{\in}{\Omega},\eqno(4.21)$$
where $K^{\star}$ is the adjoint of $K$. Hence, the matrix 
$K\bigl({\theta}(x_{0})\bigr)$ is symmetric for 
all points $x_{0}$ in ${\Omega}$. This is equivalent to the required 
result.
\vskip 0.5cm
\centerline {\bf 5. Fluctuating Currents, Chaoticity and the Onsager-
Machlup Process}
\vskip 0.3cm
{\bf 5.1. A Preliminary Observation.} We now aim to extend the 
stochastic process ${\xi}$ so as to include the currents associated 
with these fluctuations. To this end we recall that, under the 
assumptions of Props. 3.1 and 3.3, ${\xi}_{t}^{(N)}$ converges a 
classical process ${\xi}$, indexed by ${\cal 
D}^{{\prime}m}({\Omega}){\times}
{\bf R}$, with ${\xi}_{t}(f)$ continuous in $f$ and measurable in $t$. 
We shall now argue that, by contrast, ${\eta}^{(N)}$ cannot converge to 
a process ${\eta}$ possessing the corresponding continuity and 
measurability properties. To show this, we suppose that the correlation 
functions for ${\eta}^{(N)}$ converge to those of a process ${\eta}$, 
indexed by ${\cal D}_{V}^{{\prime}m}({\Omega}){\times}{\bf R}$. Then, 
since ${\cal L}^{\star}$ is the generator of $T^{\star}({\bf R}_{+})$, 
it follows from Eqs. (3.15), (3.20), (4.1), (4.5) and (4.6) that
$$\int_{0}^{t}ds_{1}\int_{0}^{t}ds_{2}E\bigl({\eta}_{s_{1}}({\nabla}f)
{\eta}_{s_{2}}({\nabla}f)\bigr)=E\bigl([{\xi}_{t}(f)-
{\xi}(f)]^{2}\bigr)=$$
$$2E\bigl({\xi}(f)[{\xi}(f)-{\xi}(T_{t}^{\star}f)]\bigr)=-
2\int_{0}^{t}dsW_{S}(f,T_{t}^{\star}{\cal L}^{\star}f)) \ 
{\forall}f{\in}{\cal D}^{m}({\Omega}), \ t{\in}{\bf R}_{+}.$$
Now the r.h.s. of this equation is $O(t)$, whereas the l.h.s. would be 
$O(t^{2})$ if $E\bigl({\eta}_{s_{1}}(g){\eta}_{s_{2}}(g)\bigr)$ were 
continuous in $g$ and measurable with respect to $s_{1}$ and $s_{2}$. 
Hence, we cannot assume that ${\eta}^{(N)}$ converges to a process 
${\eta}$ that possesses these continuity and measurability properties.
\vskip 0.3cm
{\bf 5.2. The Processes ${\zeta}$ and ${\eta}$.} In view of this 
observation, we proceed somewhat differently, starting with the 
definition  
$${\zeta}_{t,s}^{(N)}(g):=\int_{s}^{t}du{\eta}_{u}^{(N)}(g) \
{\forall} \ g{\in}{\cal D}_{V}^{m}({\Omega}), \ t,s{\in}
{\bf R}.\eqno(5.1)$$
We assume that the cyclic vector ${\Phi}^{(N)}$ lies in the domain of 
all monomials in the operators ${\xi}_{u}^{(N)}(f)$ and 
${\zeta}_{t,s}^{(N)}(g)$ as $f$ and $g$ run through ${\cal 
D}^{m}({\Omega})$ and ${\cal D}_{V}^{m}({\Omega})$, respectively, and 
$t,s$ and $u$ run through 
${\bf R}$. We further assume that the correlation functions given by 
the expectation values of these monomials for the vector state 
${\Phi}^{(N)}$ are continuous in their spatial test functions and time 
variables, that they converge pointwise to definite limits as 
$N{\rightarrow}{\infty}$, and that these limits satisfy the canonical 
counterparts to the assumptions of Props. (3.1) and (3.3). It then 
follows, by analogy with the arguments of Section 3, that the quantum 
process $({\xi}^{(N)},{\zeta}^{(N)})$ converges to a classical one, 
$({\xi},{\zeta})$, whose two components are indexed by 
${\cal D}^{m}({\Omega}){\times}{\bf R}$ and
${\cal D}_{V}^{m}({\Omega}){\times}{\bf R}^{2}$, respectively, and are 
continuous with respect to their spatial test functions and measurable 
with respect to their time variables.
\vskip 0.2cm
In view of Eq. (5.1) and the fact that the process ${\zeta}$ is the 
limiting form of ${\zeta}^{(N)}$ as $N{\rightarrow}{\infty}$, we term 
${\zeta}$ the {\it time-integrated current}. We note that since by Eq. 
(5.1),
$${\zeta}_{t,s}^{(N)}{\equiv}{\zeta}_{t,u}^{(N)}+{\zeta}_{u,s}^{(N)}  \ 
{\rm and} \ {\zeta}_{t,t}^{(N)}{\equiv}0,$$
it follows that, correspondingly,
$${\zeta}_{t,s}{\equiv}{\zeta}_{t,u}+{\zeta}_{u,s} \ 
{\rm and} \ {\zeta}_{t,t}{\equiv}0,\eqno(5.2)$$
Further, by Eqs. (3.15) and (5.1),
$${\xi}_{t}^{(N)}(f)-{\xi}_{s}^{(N)}(f)={\zeta}_{t,s}^{(N)}({\nabla}f) 
\ {\forall} \ f{\in}{\cal D}^{m}({\Omega}), \ t,s{\in}{\bf R},$$
and hence, correspondingly, 
$${\xi}_{t}(f)-{\xi}_{s}(f)={\zeta}_{t,s}({\nabla}f) \ 
{\forall} \ f{\in}{\cal D}^{m}({\Omega}), \ t,s{\in}{\bf R},
\eqno(5.3)$$
which is just the local conservation law for ${\xi}$.
\vskip 0.3cm
{\bf 5.3. Extension of the Regression Hypothesis: Secular and 
Stochastic Currents.} By Eq. (1.8), the increment ${\delta}j_{t}$ in 
the phenomenological current due to a perturbation ${\delta}q_{t}$ of 
the field $q_{t}$ is ${\cal K}{\delta}q_{t}$. Correspondingly, by way 
of extending the regression hypothesis of Section 3, we designate the 
secular part of the time-integrated fluctuation current ${\zeta}_{t,s}$ 
to be 
$${\zeta}_{t,s}^{sec}:=\int_{s}^{t}du{\cal K}{\xi}_{u},\eqno(5.4)$$ 
where ${\cal K}$, defined formally by Eq. (1.8), may now be interpreted 
as a mapping from ${\cal D}^{{\prime}m}({\Omega})$
into ${\cal D}_{V}^{{\prime}m}({\Omega})$. We define the time-
integrated stochastic current to be the residual part of 
${\zeta}_{t,s}$, namely 
$${\tilde {\zeta}}_{t,s}={\zeta}_{t,s}-{\zeta}_{t,s}^{sec},$$
i.e., by Eq. (5.4),
$${\tilde {\zeta}}_{t,s}={\zeta}_{t,s}-
\int_{s}^{t}du{\cal K}{\xi}_{u}.\eqno(5.5)$$
In view of this formula, we may re-express the local conservation law 
(5.3) in the form
$${\xi}_{t}(f)-{\xi}_{s}(f)=\int_{s}^{t}du{\xi}_{u}
({\cal K}^{\star}{\nabla}f)+
{\tilde {\zeta}}_{t,s}({\nabla}f),$$  
or equivalently, since Eqs. (1.9) and (3.15) imply that 
${\nabla}.{\cal K}=-{\cal L}$,
$${\xi}_{t}(f)-{\xi}_{s}(f)=\int_{s}^{t}du{\xi}_{u}({\cal L}^{\star}f)+
w_{t,s}(f) \ {\forall} \ f{\in}{\cal D}^{m}({\Omega}), \ t,s{\in}
{\bf R},\eqno(5.6)$$
where
$$w_{t,s}(f):={\tilde {\zeta}}_{t,s}({\nabla}f)) \ {\forall} \ 
f{\in}{\cal D}^{m}({\Omega}), \ t,s{\in}{\bf R}.\eqno(5.7)$$
Further, since, by Eqs. (5.2) and (5.7),
$$w_{t,s}{\equiv}w_{t,u}+w_{u,s} \ {\rm and} \ 
w_{t,t}{\equiv}0,\eqno(5.8)$$
Eq. (5.6) is {\it formally} a Langevin equation. However, the condition 
for it to qualify as a {\it bona fide} Langevin equation is that $w$ 
has the temporal stochastic properties of a Wiener process. The 
following proposition, which we shall prove in Appendix A, establishes 
that its two point function does have the requisite properties. Further 
assumptions concerning the chaoticity of the time-integrated stochastic 
current ${\tilde {\zeta}}_{t}$, which will be introduced in Section 
5.4, then lead to a picture in which $w$ is indeed a fully fledged 
Wiener process.
\vskip 0.3cm
{\bf Proposition 5.1.} {\it Assuming the regression hypothesis, the 
local conservation law (5.3) and the definition of $w_{t}$, 
$$E\bigl(w_{t,s}(f){\xi}_{u}(f^{\prime})\bigr)=0 \ {\forall} \ 
t{\geq}s{\geq}u, \ f,f^{\prime}{\in}{\cal D}^{m}({\Omega})\eqno(5.9)$$
and
$$E\bigl(w_{t,s}(f)(w_{t^{\prime},s^{\prime}}(f^{\prime})\bigr)=
-\bigl[W_{S}({\cal L}^{\star}f,f')+
W_{S}(f,{\cal L}^{\star}f')\bigr]
{\vert}[s,t]{\cap}[s^{\prime},t^{\prime}]{\vert}$$
$${\forall} \ 
t,s({\leq}t),t^{\prime},s^{\prime}({\leq}t^{\prime}){\in}{\bf R}, \ 
f,f'{\in}{\cal D}^{m}({\Omega}),\eqno(5.10)$$
where the last factor represents the length of the intersection of the 
intervals $[s,t]$ and $[s^{\prime},t^{\prime}]$ and $W_{S}$ is the two-
point function defined by Eq. (4.5). Further the process $w$ is non-
trivial, i.e. $w_{t,s}$ does not vanish.}
\vskip 0.3cm
{\bf 5.4. The Chaoticity and Temporal Continuity Hypothesss.}  We 
assume that the stochastic current is chaotic in the sense that the 
space-time correlations of ${\tilde {\zeta}}_{t,s}(x)$ are of short 
range on the microscopic scale. This assumption is designed to 
represent Boltzmann\rq s hypothesis of molecular chaos, as transferred 
from the local particle velocities to the stochastic currents. Since 
$L_{N}$ tends to infinity with $N$, it signifies that the space-time 
correlations of 
${\tilde {\zeta}}_{t,s}(x)$ are of zero range on the hydrodynamic 
scale. Further, in accordance with the central limit theorem for 
fluctuation fields with short range spatial correlations [39], we 
assume that the process ${\tilde {\zeta}}$ is Gaussian. Thus, our 
chaoticity hypothesis is that
\vskip 0.3cm\noindent
{\bf (C.1)} The process ${\tilde {\zeta}}$ is Gaussian; 
\vskip 0.2cm\noindent
{\bf (C.2)} $E\bigl({\tilde {\zeta}}_{t,s}(g)
{\tilde {\zeta}}_{t^{\prime},s^{\prime}}(g^{\prime})\bigr)=0$ 
if $(s,t){\cap}(s^{\prime},t^{\prime})={\emptyset}$; and 
\vskip 0.2cm\noindent
{\bf (C.3)} $E\bigl({\tilde {\zeta}}_{t,s}(g)
{\tilde {\zeta}}_{t^{\prime},s^{\prime}}(g^{\prime})\bigr)=0$ if 
${\rm supp}(g){\cap}{\rm supp}(g^{\prime})={\emptyset}$.
\vskip 0.2cm
It follows immediately from {\bf (C.1)} that the process 
${\tilde {\zeta}}$ is completely determined by its two-point function 
$E\bigl({\tilde{\zeta}}_{t,s}(g)
{\tilde {\zeta}}_{t^{\prime},s^{\prime}}(g^{\prime})\bigr)$. In view of 
the discussion following Eq. (5.1), this is continuous with respect to 
the test functions $g$ and $g^{\prime}$ and measurable with respect to 
the time variables $t,s,t^{\prime}$ and $s^{\prime}$. We now strengthen 
this conclusion by the following continuity hypothesis to the effect 
that it is continuous with respect to the time variables. 
\vskip 0.2cm\noindent
{\bf $({\cal C})$} The two-point function 
$E\bigl({\tilde{\zeta}}_{t,s}(g)
{\tilde {\zeta}}_{t^{\prime},s^{\prime}}(g^{\prime})\bigr)$ is 
continuous with respect to the time variables 
$t,s,t^{\prime},s^{\prime}$.
\vskip 0.2cm
The following proposition, which we shall prove in Appendix B, stems 
from a application of a key theorem of Schwartz [33, Theorem 35] to the 
process ${\tilde {\zeta}}$, subject to the assumptions {\bf (C.2)} and 
{\bf $({\cal C})$}.
\vskip 0.3cm
{\bf Proposition 5.2.} {\it Under the assumption of the hypotheses 
{\bf (C.2)}, {\bf (C.3)} and {\bf $({\cal C})$}, together with the 
condition of continuity with respect to its spatial test functions, the 
two-point function for the process ${\tilde {\zeta}}$ takes the form
$$E\bigl({\tilde{\zeta}}_{t,s}(g)
{\tilde {\zeta}}_{t^{\prime},s^{\prime}}(g^{\prime})\bigr)=
{\Gamma}(g,g^{\prime}){\vert}[s,t]{\cap}[s^{\prime},t^{\prime}]{\vert} 
\ {\forall} \ g,g^{\prime}{\in}{\cal D}_{V}^{m}({\Omega}), \ 
t,s,t^{\prime},s^{\prime}{\in}{\bf R},\eqno(5.11)$$
where ${\Gamma}{\in}{\cal D}_{V}^{{\prime}m}({\Omega}){\otimes}
{\cal D}_{V}^{{\prime}m}({\Omega})$ and ${\rm supp}{\Gamma}{\subset}
{\lbrace}(x,x^{\prime}){\in}{\Omega}^{2}{\vert}x^{\prime}=
x{\rbrace}$.}
\vskip 0.3cm
{\bf 5.5. A Local Equilibrium Condition for the Currents.} 
\vskip 0.3cm
In order to extend our local equilibrium condition to the stochastic
currents of the nonlinear diffusion process, we start by formulating 
the two point function at equilibrium for the process 
${\tilde {\zeta}}$.
\vskip 0.2cm
{\it Equilibrium Two Point Function for ${\tilde {\zeta}}$.} Assuming 
again that the field $q$ is uniform at equilibrium, we infer from Eqs. 
(1.6) and (1.7) that in this situation ${\cal L}=
{\tilde K}(q){\Delta}$, with $q$ constant. Hence, by Eqs. (3.8), (4.7), 
(5.7) and (5.10), together with the symmetry of $J(q)$, which follows 
from Eq. (2.6), 
$$E_{eq}\bigl({\tilde {\zeta}}_{t,s}({\nabla}f)
{\tilde {\zeta}}_{t^{\prime},s^{\prime}}({\nabla}f^{\prime})\bigr)=
-\Bigl[\bigl({\Delta}f,K({\theta})f^{\prime}\bigr)+
\bigl(K({\theta})f,{\Delta}f^{\prime}\bigr)\Bigr] 
{\vert}[s,t]{\cap}[s^{\prime},t^{\prime}]{\vert},$$
which, by Eq. (4.16), is equivalent to the following formula for the 
unsmeared two-point function for ${\tilde {\zeta}}$.
$${{\partial}^{2}\over {\partial}x_{\mu}{\partial}x_{\nu}'}
E_{eq}\bigl({\tilde {\zeta}}_{t,s;k,{\mu}}(x)
{\tilde {\zeta}}_{t^{\prime},s^{\prime};l{\nu}}(x')\bigr)=
-2K_{kl}({\theta}){\Delta}{\delta}(x-x') 
{\vert}[s,t]{\cap}[s^{\prime},t^{\prime}]{\vert},
\eqno(5.12)$$
where ${\tilde {\zeta}}_{t,s;k,{\mu}}$ is the ${\mu}$\rq th spatial 
component of the $k$\rq th component of the field 
${\tilde {\zeta}}_{t,s}=({\tilde {\zeta}}_{t,s;1}, ..\ , 
{\tilde {\zeta}}_{t,s;m})$ and the summation convention is employed for 
the indices ${\mu}$ and ${\nu}$. Recalling now our assumption, at the 
start of Section 2, that the interactions are translationally and 
rotationally invariant, we assume that the corresponding symmetries are
unbroken in the pure equilibrium phase and thus that the process
${\tilde {\zeta}}$ is invariant under the space translations and
rotations that are implemented within the confines of ${\Omega}$.
We remark here that the limitation in Euclidean symmetry imposed
by the boundedness of ${\Omega}$ is not serious from the physical
standpoint, since ${\Omega}$ is an open subset of $X$ and so any
point of it, as viewed in the microscopic picture, is infinitely
far from the boundary of ${\Sigma}$. 
\vskip 0.2cm
Assuming then that the equilibrium two-point function for 
${\tilde {\zeta}}$ is invariant under space translations and rotations, 
we may express it in the form 
$$E_{eq}\bigl({\tilde {\zeta}}_{t,s;k,{\mu}}(x)
{\tilde {\zeta}}_{t^{\prime},s^{\prime};l,{\nu}}(x')\bigr)=
S_{kl}(x-x'){\delta}_{{\mu}{\nu}}
{\vert}[s,t]{\cap}[s^{\prime},t^{\prime}]{\vert},\eqno(5.13)$$
where $S_{kl}{\in}{\cal D}^{\prime}({\Omega})$. It follows from this 
formula that Eq. (5.12) reduces to the
following differential equation for $S_{kl}$.
$${\Delta}S_{kl}(x)=2K_{kl}({\theta})
{\Delta}{\delta}(x).\eqno(5.14)$$
Further, by condition {\bf (C.3)} and Eq. (5.13), the distribution 
$S_{kl}$ has support at the origin, and therefore [33, Theorem 35] 
$S_{kl}(x-x',t)$ is a finite linear combination of ${\delta}(x-
x^{\prime})$ and its derivatives. Hence the only admissible solution of 
Eq. (5.14) is
$$S_{kl}(x)=2K_{kl}({\theta}){\delta}(x)$$
and therefore, by Eq. (5.13), the equilibrium two-point function for 
${\tilde {\zeta}}$ is given by the formula
$$E_{eq}\bigl({\tilde {\zeta}}_{t,s;k,{\mu}}(x)
{\tilde {\zeta}}_{t^{\prime},s^{\prime};l,{\nu}}(x')\bigr)=
2K_{kl}({\theta}){\delta}(x-x'){\delta}_{{\mu}{\nu}}
{\vert}[s,t]{\cap}[s^{\prime},t^{\prime}]{\vert} \ .\eqno(5.15)$$
Equivalently, the equilibrium two-point function for the smeared
field ${\tilde {\zeta}}_{t,s}(g)$ takes the form 
$$E_{eq}\bigl({\tilde {\zeta}}_{t,s}(g)
{\tilde {\zeta}}_{t^{\prime},s^{\prime}}(g')\bigr)=
2\bigl(g,K({\theta})g')_{V}
{\vert}[s,t]{\cap}[s^{\prime},t^{\prime}]{\vert}$$
$$ {\forall} \ g,g'{\in}{\cal D}_{V}^{m}({\Omega}),
\ t,s,t^{\prime},s^{\prime}{\in}{\bf R},\eqno(5.16)$$
where $(.)_{V}$ is the inner product in 
${\cal D}_{V}^{m}({\Omega})$ defined by the formula
$$(g,g')_{V}={\sum}_{k=1}^{m}\int_{\Omega}dxg(x).g'(x) \
{\forall} \ g,g'{\in}{\cal D}_{V}^{m}({\Omega}){\equiv}$$
$${\sum}_{k=1}^{m}{\sum}_{{\mu}=1}^{d}\int_{\Omega}dx
g_{k,{\mu}}(x)g_{k,{\mu}}^{\prime}(x),\eqno(5.17)$$
and where $g_{k,{\mu}}$ is the ${\mu}$\rq th spatial component of 
$g_{k}$.
\vskip 0.3cm
{\it Local Property of the Equilibrium Two Point Function.} We
formulate the local properties of the stochastic current 
${\tilde {\zeta}}$ along the lines employed in Section 4.2 for the
process ${\xi}$. Thus, for $(x_{0},{\epsilon}){\in}
{\Omega}{\times}{\bf R}_{+}$, and ${\epsilon}$ sufficiently
small, we define the transformation 
$g{\rightarrow}g_{x_{0},{\epsilon}}$ of 
${\cal D}_{V}^{m}({\Omega})$ by the formula
$$g_{x_{0},{\epsilon}}(x)=
{\epsilon}^{-d/2}g\bigl({\epsilon}^{-1}(x-x_{0})\bigr).
\eqno(5.18)$$
We then observe that, by Eqs. (5.17) and (5.18), the transformations 
$t{\rightarrow}{\epsilon}^{2}t, \ g{\rightarrow}g_{x_{0},{\epsilon}}$, 
of the times and test functions lead to the multiplication of the 
smeared two-point function of Eq. (5.16) by the factor 
${\epsilon}^{2}$. Thus,  
$${\epsilon}^{-2}E_{eq}
\bigl({\tilde {\zeta}}_{{\epsilon}^{2}t,{\epsilon}^{2}s}
(g_{x_{0},{\epsilon}}){\tilde 
{\zeta}}(g_{x_{0},{\epsilon}}^{\prime})\bigr)=
2\bigl(g,K({\theta})g')_{V}{\vert}[s,t]{\cap}
[s^{\prime},t^{\prime}]{\vert}$$
$${\forall} \ x_{0}{\in}{\Omega}, \ g,g'{\in}
{\cal D}_{V}^{m}({\Omega}),
\ t,s,t^{\prime},s^{\prime}{\in}{\bf R}.\eqno(5.19)$$
The local property of the two-point function for 
${\tilde {\zeta}}$ at the point $x_{0}$ is then
obtained by passing to the limiting form of this equation as
${\epsilon}{\rightarrow}0$.
\vskip 0.3cm
{\it Local Equilibrium Property for the Stochastic Current in the
Nonequilibrium Steady State.} In view of the last observation,
we assume that, in the nonequilibrium steady state, the process
${\tilde {\zeta}}$ enjoys the local equilibrium property obtained
by passing to the limit ${\epsilon}{\rightarrow}0$ and replacing
$E_{eq}$ and ${\theta}$ by $E$ and ${\theta}(x_{0})$,
respectively, in Eq. (5.19). Thus we assume that 
$${\rm lim}_{{\epsilon}{\rightarrow}0}{\epsilon}^{-2}
E\bigl({\tilde {\zeta}}_{{\epsilon}^{2}t,{\epsilon}^{2}s}
(g_{x_{0},{\epsilon}})
{\tilde {\zeta}}_{{\epsilon}^{2}t^{\prime},{\epsilon}^{2}s^{\prime}}
(g_{x_{0},{\epsilon}}^{\prime})\bigr)=
2\bigl(g,K({\theta}(x_{0}))g^{\prime})_{V}
{\vert}[s,t]{\cap}[s^{\prime},t^{\prime}]{\vert}$$
$${\forall} \ x_{0}{\in}{\Omega}, \ g,g^{\prime}{\in}
{\cal D}_{V}^{m}({\Omega}), \ 
t,s({\leq}t),t^{\prime},s^{\prime}({\leq}t^{\prime}){\in}
{\bf R}.\eqno(5.20)$$
This is our local equilibrium condition for the stochastic current. 

\vskip 0.3cm
{\bf 5.6. Explicit Form of the Two Point Function for ${\tilde 
{\zeta}}$.} By Prop. 5.2, this function is determined by the functional 
${\Gamma}$, which by Eqs. (5.11) and (5.20), possesses the following 
local equilibrium property.
$${\rm lim}_{{\epsilon}{\downarrow}0}{\Gamma}(g_{x_{0},{\epsilon}}, 
g_{x_{0},{\epsilon}}^{\prime})=
2\bigl(g,K({\theta}(x_{0}))g^{\prime})_{V}
{\forall} \ x_{0}{\in}{\Omega}, \ g,g^{\prime}{\in}
{\cal D}_{V}^{m}({\Omega}).\eqno(5.21)$$
The following proposition, which will be proved in Appendix C, provides 
an explicit formula for the functional ${\Gamma}$, which stems from a 
combination of the chaoticity condition {\bf (C.3)} and the local 
equilibrium condition (5.21). 
\vskip 0.3cm
{\bf Proposition 5.3.} {\it Under the previous assumptions, together 
with the local equilibrium condition of (5.21), ${\Gamma}$ is given by 
the formula
$${\Gamma}(g,g^{\prime})=2(g,K_{\theta}g')_{V} \ {\forall} \ g,g'{\in}
{\cal D}_{V}^{m}({\Omega}),\eqno(5.22)$$
where $K_{\theta}$ is the matrix-valued operator
$K{\circ}{\theta}$ in ${\cal D}_{V}^{m}({\Omega})$, i.e.
$$K_{\theta}(x)=K\bigl({\theta}(x)\bigr).\eqno(5.23)$$}
\vskip 0.3cm
The following corollary is an immediate consequence of this proposition 
and Prop. 5.2.
\vskip 0.3cm
{\bf Corollary 5.4.} {\it Under the same assumptions, the two-point 
function of the stationary process ${\tilde {\zeta}}$ is given by the 
formula 
$$E\bigl({\tilde {\zeta}}_{t,s}(g)
{\tilde {\zeta}}_{t^{\prime},s^{\prime}}(g')\bigr)=
2(g,K_{\theta}g')_{V}{\vert}[s,t]{\cap}[s^{\prime},t^{\prime}]{\vert}$$
$${\forall} \ g,g'{\in}
{\cal D}_{V}^{m}({\Omega}), \ 
t,s({\leq}t),t',s^{\prime}({\leq}t^{\prime})
{\in}{\bf R}.\eqno(5.24)$$} 
\vskip 0.3cm
{\bf 5.7. The Generalized Onsager-Machlup Process ${\xi}$.} 
\vskip 0.3cm
It now follows immediately from Cor. 5.4 and Eq. (5.7) that
$$E\bigl(w_{t,s}(f)w_{t^{\prime},s^{\prime}}(f^{\prime})\bigr)=
2({\nabla}f,K_{\theta}{\nabla}f^{\prime})_{V}
{\vert}[s,t]{\cap}[s^{\prime},t^{\prime}]{\vert}$$
$${\forall} \ f,f'{\in}{\cal D}^{m}({\Omega}), \ 
t,s,t^{\prime},s^{\prime}{\in}
{\bf R}.\eqno(5.25)$$ 
Hence, by the chaotic hypothesis {\bf (C.1)} and Eqs. (5.7) and (5.25), 
$w$ is a generalized Wiener process. Further, on re-expressing Eq. 
(5.6) in the form
$$d{\xi}_{t}={\cal L}{\xi}_{t}dt+dw_{t,s},\eqno(5.26)$$
we see that, in view of the additive property (5.8) of $w$,
the fluctuation field ${\xi}$ executes a {it generalized Onsager-
Machlup process}; while Eq. (5.25) signifies that the two-point 
function for $w$ corresponds precisely to that assumed for the 
stochastic force in Landau's fluctuation hydrodynamics [18]. 
\vskip 0.2cm
In order to derive the properties of the process ${\xi}$ from those of 
$w$, we note that, since ${\cal L}$ is the generator of 
$T({\bf R}_{+})$, the solution of the Langevin equation (5.26) is given 
by the formula
$${\xi}_{t}=T_{t-s}{\xi}_{s}+\int_{s}^{t}T_{t-u}dw_{u,s} \ 
{\forall} \ t,s({\leq}t){\in}{\bf R},\eqno(5.27)$$
or equivalently, 
$${\xi}_{t}(f)={\xi}_{s}(T_{t-s}^{\star}f)+\int_{s}^{t}dw_{u,s}
(T_{t-u}^{\star}f) \ {\forall} \ f{\in}{\cal D}^{m}({\Omega}), \ 
T,s({\leq}t){\in}{\bf R}_{+}.\eqno(5.28)$$ 
\vskip 0.2cm
The following proposition, which we shall prove in Appendix D, is a 
natural generalization of standard properties of the Brownian motion of 
a single particle that ensue from the Langevin equation governing its 
velocity (cf. [36]).
\vskip 0.3cm
{\bf Proposition 5.5.} {\it Under the above assumptions,
\vskip 0.2cm\noindent
(1) ${\xi}$ is a Gaussian Markov process, and 
\vskip 0.2cm\noindent
(2) the fields $w_{t,s}$ and ${\xi}_{u}$ are statistically independent 
of one another if $s$ and $t$ are greater than or equal to $u$.}
\vskip 0.3cm
{\bf Comment.} It follows from this proposition and Eqs. (4.5) and 
(4.6) that the process ${\xi}$ is completely determined by the forms of 
the semigroup $T^{\star}({\bf R}_{+})$ and the distribution $W_{S}$.
\vskip 0.5cm
\centerline {\bf 6. Long Range Spatial Correlations of the ${\xi}$-
Process} 
\vskip 0.3cm
{\bf 6.1. The Static Two-Point Function for ${\xi}$.} By Eq. (4.5), the 
unsmeared form of the  ${\cal D}^{{\prime}m}({\Omega}){\otimes}
{\cal D}^{{\prime}m}({\Omega})$-class 
distribution $W_{S}$ is given by the formula
$$W_{S}(x,x^{\prime})=E\bigl({\xi}(x){\otimes}{\xi}(x^{\prime})\bigr).
\eqno(6.1)$$
\vskip 0.2cm
The  following Proposition provides an explicit formula for $W_{S}$, as 
well as a differential equation for this distribution in terms of the 
semigroup $T^{\star}({\bf R}_{+})$, and the transport function 
$K_{\theta}$. 
\vskip 0.3cm
{\bf Proposition 6.1.} {\it Under the above assumptions,
$$W_{S}(f,f^{\prime})=2\int_{0}^{\infty}dt\bigl({\nabla}T_{t}^{\star}f,
K_{\theta}{\nabla}T_{t}^{\star}f^{\prime}\bigr)_{V}
\ {\forall} f,f^{\prime}{\in}{\cal D}^{m}({\Omega})\eqno(6.2)$$
and, further, the generalized function $W_{S}(x,x^{\prime})$ satisfies 
the equation 
$$[{\cal L}{\otimes}I+I{\otimes}{\cal L}^{\prime}]W_{S}(x,x^{\prime})=
2{\nabla}.\bigl(K_{\theta}(x){\nabla}{\delta}(x-
x^{\prime})\bigr),\eqno(6.3)$$
where ${\cal L}^{\prime}$ is the version of ${\cal L}$ that acts on 
functions of $x^{\prime}$. }
\vskip 0.3cm
{\bf Proof.} By Eq. (4.5) and the stationarity of the ${\xi}$-process,
$$W_{S}(f,f^{\prime})=E\bigl({\xi}_{t}(f){\xi}_{t}(f^{\prime})\bigr) \ 
{\forall} 
\ f,f^{\prime}{\in}{\cal D}^{m}({\Omega}), \  t{\in}{\bf R}_{+}$$
and therefore, by Eq. (5.26), 
$$W_{S}(f,f^{\prime})=E\bigl({\xi}(T_{t}^{\star}f)
{\xi}(T_{t}^{\star}f^{\prime})\bigr)+
\int_{0}^{t}E\bigl({\xi}(T_{t}^{\star}f)dw_{u,0}
(T_{t-u}^{\star}f^{\prime})\bigr)+$$ 
$$\int_{0}^{t}E\bigl({\xi}(T_{t}^{\star}f^{\prime})
dw_{u,0}(T_{t-u}^{\star}f)\bigr)+
\int_{0}^{t}\int_{0}^{t}E\bigl(dw_{u,0}(T_{t-u}^{\star}f)
dw_{u^{\prime},0}(T_{t-u^{\prime}}^{\star}f^{\prime})\bigr) $$
$$ \ {\forall} \ f,f^{\prime}{\in}
{\cal D}^{m}({\Omega}), \  t{\in}{\bf R}_{+}.\eqno(6.4)$$
Now, by the dissipativity condition (3.12), the first term on the 
r.h.s. of this 
equation vanishes in the limit $t{\rightarrow}{\infty}$, while by Eq. 
(5.9), the second and third terms there vanish. Hence, it follows from 
Eq. (6.4) that
$$W_{S}(f,f^{\prime})={\rm lim}_{t\to\infty}\int_{0}^{t}\int_{0}^{t}
E\bigl(dw_{u,0}(T_{t-u}^{\star}f)dw_{u^{\prime},0}
(T_{t-u^{\prime}}^{\star}f^{\prime})\bigr) \ {\forall} \ 
f,f^{\prime}{\in}
{\cal D}^{m}({\Omega}) .\eqno(6.5)$$
Further, by Eq. (5.25),
$$E\bigl(dw_{u,0}(f)dw_{u^{\prime},0}(f^{\prime})\bigr)=
2({\nabla}f,K_{\theta}{\nabla}f^{\prime})_{V}{\delta}(u-
u^{\prime})dudu^{\prime}$$
and consequently Eq. (6.5) reduces to the form
$$W_{S}(f,f^{\prime})={\rm lim}_{t\to\infty}
2\int_{0}^{t}du({\nabla}T_{t-u}^{\star}f,{\nabla}
T_{t-u}^{\star}f^{\prime})_{V}{\equiv}
2\int_{0}^{t}du({\nabla}T_{u}^{\star}f,{\nabla}
T_{u}^{\star}f^{\prime})_{V},$$ 
which is equivalent to the required formula (6.2).
\vskip 0.2cm
Further, since ${\cal L}^{\star}$ is the generator of 
$T^{\star}({\bf R}_{+})$, 
it follows  from Eq. (6.2) that
$$W_{S}({\cal L}^{\star}f,f^{\prime})+
W_{S}(f,{\cal L}^{\star}f^{\prime})=
2\int_{0}^{\infty}dt{d\over dt}
({\nabla}T_{t}^{\star}f,
K_{\theta}{\nabla}T_{t}^{\star}f^{\prime})_{V}$$
and consequently, by the dissipativity condition (3.12),
$$W_{S}({\cal L}^{\star}f,f^{\prime})+
W_{S}(f,{\cal L}^{\star}f^{\prime})=
-2({\nabla}f,K_{\theta}{\nabla}f^{\prime})_{V} \ {\forall} \ 
f,f^{\prime}{\in}
{\cal D}^{m}({\Omega}),$$
which, by Eq. (6.1), is equivalent to the required formula (6.3). 
\vskip 0.3cm
{\bf 6.2. Long Range Spatial Correlations.} In order to provide a 
precise characterization of long range correlations, we 
first recall that the ratio of the macroscopic length scale to the 
microscopic one is infinite. Consequently, correlations of finite range  
on the microscopic scale are of zero range on the macroscopic one. 
Accordingly, we term the range of correlations \lq short\rq\ or \lq 
long\rq\  according to whether or 
not it reduces to zero in the macroscopic picture. Thus our condition 
for long 
range spatial correlations for the ${\xi}$-field is simply that the 
support
of the distribution $W_{S}$ does {\it not} lie in the domain
${\lbrace}(x,x^{\prime}){\in}{\Omega}^{2}{\vert}x=x^{\prime}{\rbrace}$. 
The
following proposition establishes that the spatial correlations
of ${\xi}$ for the nonlinear diffusion process are generically of long 
range.
\vskip 0.3cm
{\bf Proposition 6.2.} {\it Let ${\Phi}_{q}$ be the $m$-by-$m$
matrix-valued function on ${\Omega}$ defined by
the formula
$${\Phi}_{q}(x)={\Delta}K_{\theta}(x)+
{\nabla}.{\Psi}_{q}(x),\eqno(6.6)$$
where
$${\Psi}_{q;kl}(q;x)={\sum}_{k^{\prime},l^{\prime}=1}^{m}
\Bigl[{{\partial}\over {\partial}q_{l^{\prime}}(x)}
{\tilde K}_{kk^{\prime}}\bigl(q(x)\bigr)\Bigr]
\bigl[J_{l^{\prime}l}\bigl(q(x)\bigr){\nabla}q_{k^{\prime}}(x)-
J_{k^{\prime}l}\bigl(q(x)\bigr){\nabla}q_{l^{\prime}}(x)\bigr].
\eqno(6.7)$$ 
Then under the above assumptions, a sufficient
condition for the spatial correlations of ${\xi}$ to be of long
range is that either ${\Phi}_{q}$ does not vanish or that the matrix 
${\Psi}_{q}$ is symmetric.}
\vskip 0.3cm
{\bf Comments.} (1) The Proposition establishes that
the correlations are generically of long range, since the specified
conditions on ${\Phi}_{q}$ and ${\Psi}_{q}$ can be satified only for
special relationships between the functions ${\tilde K}{\circ}q$
and $s{\circ}q$; and these are generally independent of one another, 
since $s$ and ${\tilde K}$ govern the equilibrium and transport 
properties, respectively, of ${\Sigma}$. By contrast, the corresponding
correlations for equilibrium states are generically of short
range, except at critical points. A treatment of critical
equilibrium correlations of fluctuation observables is provided
by Ref. [40].
\vskip 0.2cm 
(2) In the particular case of the symmetric
exclusion process [9-11], $n=1, \ d=1, \ {\tilde K}(q)=1, \
s(q)=-q{\ln}q-
(1-q){\ln}(1-q)$ and $q(x)=a+b.x$, where $a$ and $b \ ({\neq}0)$
are constants. Thus, in this case, it follows from Eqs. (1.6), (2.6),
(6.6) and (6.7) that ${\Psi}_{q}=0$ and ${\Phi}_{q}(x)=-2b^{2}{\neq}0$. 
Hence, long range correlations prevail in this model, in accordance 
with the results obtained by its explicit solution in Refs. [9-11]. 
\vskip 0.3cm
{\bf Proof of Prop. 6.2.} Suppose that the static spatial correlations 
of ${\xi}$ are not of long range, i.e. that the support of the 
distribution $W_{S}$ lies in the domain ${\lbrace}(x,x^{\prime}){\in}
{\Omega}^{2}{\vert}x^{\prime}=x{\rbrace}$. Then it follows from this 
supposition and the local equilibrium condition (4.14), by precise 
analogy of the derivation of Eq. (5.24) from corresponding conditions 
of zero range correlations and local equilibrium for the process 
${\tilde {\zeta}}$, that
$$W_{S}(x,x^{\prime})=J_{q}(x){\delta}(x-x^{\prime}),\eqno(6.8)$$ 
where
$$J_{q}(x):=J\bigl(q(x)\bigr).\eqno(6.9)$$
Hence, by Eqs. (1.10), (3.8) and (6.7)-(6.9),
$$({\cal L}{\otimes}I)W(x,x^{\prime})=
{\Delta}[K_{\theta}(x){\delta}(x-x^{\prime})]+
{\nabla}.[{\Psi}_{q}(x){\delta}(x-x^{\prime})].\eqno(6.10)$$
Further, by Eq. (6.1),
$$(I{\otimes}{\cal L}^{\prime})W(x,x^{\prime})=
[({\cal L}^{\prime}{\otimes}I)W(x^{\prime},x)]^{tr},$$
where the superscript $tr$ denotes transpose, and therefore, by Eq. 
(6.10),
$$(I{\otimes}{\cal L}^{\prime})W(x,x^{\prime})=
{\Delta}^{\prime}[K_{\theta}(x^{\prime}-x){\delta}(x^{\prime}-x)]^{tr}+
{\nabla}^{\prime}.[{\Psi}_{q}(x^{\prime}){\delta}(x^{\prime}-x)]^{tr},
\eqno(6.11)$$
where ${\Delta}^{\prime}$ and ${\nabla}^{\prime}$ are the versions of 
${\Delta}$ and ${\nabla}$, respectively, that act on functions of 
$x^{\prime}$. Consequently, since $K_{\theta}$ is symmetric, by Eqs. 
(4.16) and (5.23), it follows from Eqs. (6.6), (6.10) and (6.11) that
$$[{\cal L}{\otimes}I+I{\otimes}
{\cal L}^{\prime}]W_{S}(x,x^{\prime})=$$
$$2{\nabla}.\bigl(K_{\theta}(x){\nabla}{\delta}(x-x^{\prime})\bigr)+
{\Phi}_{q}(x){\delta}(x-x^{\prime})+[{\Psi}_{q}(x)-{\Psi}_{q}^{tr}(x)].
{\nabla}{\delta}(x-x^{\prime}).\eqno(6.12)$$
On comparing this equation with Eq. (6.3), we see that
$${\Phi}_{q}(x){\delta}(x-x^{\prime})+[{\Psi}_{q}(x)-
{\Psi}_{q}^{tr}(x)].{\nabla}{\delta}(x-x^{\prime})=0,$$
i.e. that ${\Phi}_{q}$ vanishes and that ${\Psi}_{q}$ is symmetric. 
These, then, are conditions that ensue from the assumption of short 
range correlations of the ${\xi}$-process. We conclude, therefore, that 
the violation of either of these conditions signifies that the 
correlations are of long range.
\vskip 0.5cm
\centerline {\bf 7. Concluding Remarks.}
\vskip 0.3cm
We have proposed a macrostatistical treatment of nonequilibrium stady 
states of 
quantum systems that is centred on the fluctuations of their 
hydrodynamical 
variables. The key physical assumptions on which this treatment is 
based are 
\vskip 0.2cm\noindent
(a) the regression hypothesis for the hydrodynamic fluctuation field 
${\xi}$;
\vskip 0.2cm\noindent
(b) the chaoticity of the associated currents, as represented by their 
time 
integrals ${\zeta}_{t,s}$;
\vskip 0.2cm\noindent
(c) the local equilibrium conditions on the stochastic process 
comprising 
${\xi}$ and ${\zeta}$;
\vskip 0.2cm\noindent
(d) the space-time scale invariance of the phenomenological
equation of motion (1.4), as exemplified by the case of nonlinear
diffusions; and 
\vskip 0.2cm\noindent
(e) the invariance of the quantum field ${\hat q}$, and correspondingly 
of the classical field ${\xi}$, under time reversals.
\vskip 0.2cm\noindent
On the basis of these assumptions and certain technical ones, we have 
obtained a picture that provides natural generalizations of the Onsager 
reciprocity relations and the Onsager-Machlup fluctuation process to 
nonequilibrium steady states, together with a demonstration that the 
spatial correlations of the hydrodynamical variables are generically of 
long range in these states. Furthermore this picture is expressed 
exclusively in terms of the phenomenological functions representing the 
equilibrium entropy, $s(q)$, the transport coefficients $K({\theta})$ 
and the hydrodynamical boundary conditions. This may easily be seen 
from the comment at the end of Section 5, together with Eqs. (1.10), 
(3.8) and (6.2) and the fact that the semigroup $T({\bf R})$ is 
completely determined by its generator ${\cal L}$.
\vskip 0.2cm
Let us now discuss the assumptions (a)-(e) a little further. In our 
view, for reasons expressed in Sections 4.1, 4.2 and 5.4, the first 
three of these seem natural from the physical standpoint, though they 
are very hard to prove in concrete cases. On the other hand, it is 
clear that assumptions (d) and (e) are not universally valid: for 
example, they both fail in the important case of Navier-Stokes 
hydrodynamics. Consequently, it is of interest to consider how the 
macrostatistical picture presented here might be extended to situations 
where (d) and (e) are replaced by weaker assumptions.
\vskip 0.2cm
In fact, the weakening of (e) provides no serious problems, since
the locally conserved fields of continuum mechanics are generally 
either even or odd with respect to time reversals [41]. Accordingly, 
we replace (e) by the assumption that each of the quantum fields 
${\hat q}_{j}$ has either even or odd parity with respect to time 
reversals, 
i.e. that
$${\tau}{\hat q}_{j}(x)=R_{j}{\hat q}_{j}(x), \  R_{j}={\pm}1,
\ j=1,. \ .,n,\eqno(7.1)$$
where again ${\tau}$ is the time-reversal antiautomorphism. This
weakened assumption then leads to the nonlinear version of
Casimir's extension [41] of Onsager's theory, wherein Eq. (4.16)
is modified to the formula
$$K_{kl}\bigl({\theta}(x_{0})\bigr)=
R_{k}R_{l}K_{lk}\bigl(R{\theta}(x_{0})\bigr),\eqno(7.2)$$
where
$$R({\theta}):=\bigl(R_{1}{\theta}_{1},. \
.,R_{n}{\theta}_{n}\bigr).\eqno(7.3)$$
Similarly, the modification of assumption (e) to the form given
by Eq. (7.1) presents no serious problems for the other issues
treated here.  
\vskip 0.2cm
On the other hand, there does not appear to be any natural
generalisation of the scaling assumption (d), which lay behind the 
interdependence of the ratios of the macroscopic to microscopic scales 
for 
distance and time, the former ratio being $L_{N}$ and the latter 
$L_{N}^{2}$ (or 
more generally $L_{N}^{k}$). Moreover, one sees from Eqs. (2.15) and 
(3.13) that 
this interdependence was essential to the limit procedures of Eqs. 
(3.1) and 
(3.20). Nevertheless it does not appear to be essential to the key 
physical 
ideas that
\vskip 0.2cm\noindent
(i) the ratios of the macroscopic to microscopic 
scales for both distance and time are extremely large, and 
\vskip 0.2cm\noindent
(ii) the currents 
associated with the locally conserved quantum fields satisfy the 
chaoticity assumption of Section 5.4, whereby the space-time 
correlations of 
their 
fluctuations decay within microscopic distances and times. 
\vskip 0.2cm\noindent
Since such chaoticity does not necessarily require any interdependence 
of the 
ratios of the macroscopic to microscopic scales for distance and time, 
it 
appears reasonable to expect that some version of the present 
macrostatistical 
model should still be applicable even in the absence of macroscopic 
space-time 
scale invariance.
\vskip 0.2cm
Thus, from the standpoint of mathematical physics, a most challenging 
question is whether the present scheme can be generalized to a setting 
which does not require the scale invariance of the macroscopic law 
(1.4). Presumably such a generalization would require a difficult 
multi-scale analysis.
\vskip 0.5cm
\centerline {\bf Appendix A: Proof of Proposition 5.1}
\vskip 0.3cm
We shall first prove Eqs. (5.9) and (5.10) and then demonstrate
the nontriviality of the process $w$.
\vskip 0.2cm
Since ${\cal L}$ is the generator of $T({\bf R}_{+})$, Eq. (5.9)
follows immediately from Eqs. (4.2) and (5.6).
\vskip 0.2cm
It then follows from Eqs. (5.6) and (5.9) that the l.h.s.
of Eq. (5.10) vanishes if the intervals $[s,t]$ and $[s',t']$ do
not intersect. Hence, in view of Eq. (5.8), the proof of Eq. (5.10) 
reduces to that of the same formula with $s=s'$ and $t=t'$ and 
$t{\geq}s$. Thus it suffices for us to prove that
$$E\bigl(w_{t,s}(f)w_{t,s}(f')\bigr)=
-\bigl(W_{S}({\cal L}^{\star}f,f')+W_{S}(f,
{\cal L}^{\star}f')\bigr){\vert}t-s{\vert}$$
$${\forall} \ t,s \ ({\leq}t) \ {\in} \
{\bf R}, \ f,f'{\in}{\cal D}^{m}({\Omega}).\eqno(A.1)$$
\vskip 0.2cm
We start by inferring from Eq. (5.6) that the l.h.s. of Eq. (A.1) is 
the
sum of the following four terms:-
$$E\bigl[\bigl({\xi}_{t}(f)-{\xi}_{s}(f)\bigr)
\bigl({\xi}_{t}(f')-{\xi}_{s}(f')\bigr)\bigr],\eqno(a)$$
$$-\int_{s}^{t}duE\bigl[\bigl({\xi}_{t}(f)-{\xi}_{s}(f)\bigr)
{\xi}_{u}({\cal L}^{\star}f')\bigr],\eqno(b)$$
$$-\int_{s}^{t}duE\bigl[{\xi}_{u}
({\cal L}^{\star}f)\bigl({\xi}_{t}(f')-{\xi}_{s}(f')\bigr)\bigr]
\eqno(c)$$
and
$$\int_{s}^{t}du\int_{s}^{t}dvE\bigl({\xi}_{u}({\cal L}^{\star}f)
{\xi}_{v}({\cal L}^{\star}f)\bigr).\eqno(d)$$
\vskip 0.2cm
Since $t{\geq}s$ and the ${\xi}$-process is stationary, it
follows from Eqs. (4.5) and (4.6) that
$${\rm Term} \ (a)=2W_{S}(f,f')-W_{S}(T_{t-s}^{\star}f,f')-
W_{S}(f,T_{t-s}^{\star}f'),\eqno(A.2)$$
$${\rm Term} \ (b)=-\int_{s}^{t}duW_{S}(T_{t-u}^{\star}f,
{\cal L}^{\star}f')+\int_{s}^{t}duW_{S}(f,T_{u-s}^{\star}
{\cal L}^{\star}f'),\eqno(A.3)$$
$${\rm Term} \ (c)=-\int_{s}^{t}duW_{S}({\cal L}^{\star}f,
T_{t-u}^{\star}f')+{\int}_{s}^{t}duW_{S}(T_{u-s}^{\star}
{\cal L}^{\star}f,f')\eqno(A.4)$$
and
$${\rm Term} \ (d)=\int_{s}^{t}du\int_{s}^{u}dv
W_{S}(T_{u-v}^{\star}{\cal L}^{\star}f,{\cal L}^{\star}f')+
\int_{s}^{t}du\int_{u}^{t}dvW_{S}({\cal
L}^{\star}f,T_{v-u}^{\star}
{\cal L}^{\star}f').\eqno(A.5)$$
\vskip 0.2cm
Since $W_{S}$ is linear in each of its arguments and since ${\cal
L}^{\star}$ is the generator of $T^{\star}({\bf R}_{+})$, it
follows that Eqs. (A.3-5) may be re-expressed in the following
forms.
$${\rm Term} \ (b)=-\int_{s}^{t}duW_{S}(T_{t-u}^{\star}f,
{\cal L}^{\star}f')+
W_{S}(f,T_{t-s}^{\star}f')-W_{S}(f,f'),\eqno(A.6)$$
$${\rm Term} \ (c)=-\int_{s}^{t}duW_{S}({\cal L}^{\star}f,
T_{t-u}^{\star}f')+W_{S}(T_{t-s}^{\star}f,f')-W_{S}(f,f')\eqno
(A.7)$$
and
$${\rm Term} \ (d)=
\int_{s}^{t}du\bigl[-W_{S}(f,{\cal
L}^{\star}f')+W_{S}(T_{u-s}^{\star}f,
{\cal L}^{\star}f')+W_{S}({\cal L}^{\star}f,T_{t-u}^{\star}f')-
W_{S}({\cal L}^{\star}f,f')\bigr].\eqno(A.8)$$
It follows now from Eqs. (A.2) and (A.6-8) that the sum of the terms 
$(a), \ (b), \ (c)$ and $(d)$, which comprises the l.h.s. of Eq. (A.1), 
is equal to the r.h.s. of that equation. This completes the proof of 
Eq. (A.1) and thus of Eq. (5.10). 
\vskip 0.2cm
Finally, we employ a {\it reductio ad absurdum} method to
establish the nontriviality of the process $w$. Thus, we assume
that $w_{t,s}$ vanishes. It then follows from Eq. (5.7) that
$$W_{S}({\cal L}^{\star}f,f')+
W_{S}(f,{\cal L}^{\star}f')=0 \ 
{\forall} \ f,f'{\in}{\cal D}^{m}({\Omega})$$
and hence that
$$W_{S}({\cal L}^{\star}T_{t}^{\star}f,T_{t}^{\star}f')+
W_{S}(T_{t}^{\star}f,{\cal L}^{\star}T_{t}^{\star}f')=0 \ 
{\forall} \ f,f'{\in}{\cal D}^{m}({\Omega}), \ t{\in}
{\bf R}_{+}.$$
Since $W_{S}$ is linear in each of its arguments and since 
${\cal L}^{\star}$ is the generator of $T^{\star}({\bf R}_{+})$,
this signifies that
$${d\over dt}W_{S}(T_{t}^{\star}f,T_{t}^{\star}f')=0$$
and therefore, since $T_{0}=I$, that 
$$W_{S}(T_{t}^{\star}f,T_{t}^{\star}f')=
W_{S}(f,f') \ {\forall} \ f,f'{\in}{\cal D}^{m}({\Omega}), \ t{\in}
{\bf R}_{+}.\eqno(A.9)$$
Moreover, by Eq. (4.5) and the dissipativity condition (3.12), the 
l.h.s. of
Eq. (A.9) vanishes in the limit $t{\rightarrow}{\infty}$. Hence Eq. 
(A.9) implies that the static two-point function $W_{S}$ vanishes. This
conflicts with the fact that, by Eqs. (4.5), (4.11) and (4.14), 
$${\rm lim}_{{\epsilon}{\downarrow}0}
W_{S}(f_{x_{0},{\epsilon}},f_{x_{0},{\epsilon}}^{\prime})=
\bigl(f,J(q(x_{0}))f'\bigr),$$
which does not vanish identically. This contradiction establishes that 
the assumption of the triviality of $w$ is untenable and thus completes 
the proof of the proposition.
\vskip 0.5cm
\centerline {\bf Appendix B: Proof of Proposition 5.2.} 
\vskip 0.3cm
We start by noting that, in view of Eq. (5.2) and condition 
{\bf (C.2)}, the proof of this proposition reduces to that of the  
formula (5.11) for the particular case where $s=s^{\prime}, \ 
t=t^{\prime}$ and $s{\leq}t$. Thus we need only prove that
$$E\bigl({\zeta}_{t,s}(g){\zeta}_{t,s}(g^{\prime})\bigr)=
{\Gamma}(g,g^{\prime})(t-s) \ {\forall} \ g,g^{\prime}
{\in}{\cal D}_{V}^{m}({\Omega}), \ t,s({\leq}t){\in}{\bf 
R},\eqno(B.1)$$
where ${\Gamma}$ is an element of 
${\cal D}_{V}^{{\prime}m}{\otimes}{\cal D}_{V}^{{\prime}m}$ with 
support in the domain  
${\lbrace}(x,x^{\prime}){\in}{\Omega}^{2}{\vert}x^{\prime}=x{\rbrace}$. 
\vskip 0.2cm
To this end, we start by defining
$$F_{g,g^{\prime}}(t,s):= 
E\bigl({\zeta}_{t,s}(g){\zeta}_{t,s}(g^{\prime})\bigr)\eqno(B.2)$$
and inferring from Eq. (5.2) and condition {\bf (C.2)} that 
$$F_{g,g^{\prime}}(t,s)=F_{g,g^{\prime}}(t,u)+F_{g,g^{\prime}}(u,s) \ 
{\rm for} \ t{\geq}u{\geq}s.\eqno(B.3)$$
Further, by Eq. (B.2) and the stationarity of the process ${\zeta}$,
$$F_{g,g^{\prime}}(t,s)=F_{g,g^{\prime}}(t+b,s+b) \ {\forall} \ b{\in}
{\bf R},$$
which signifies that $F_{g,g^{\prime}}$ may be expressed in the form
$$F_{g,g^{\prime}}(t,s)={\tilde F}_{g,g^{\prime}}(t-s) \ {\forall} \ 
s,t{\in}{\bf R},\eqno(B.4)$$ 
where, by condition ${\bf ({\cal C})}, \ {\tilde F}_{g,g^{\prime}}$ is 
a continuous function on ${\bf R}$. It follows now from Eqs. (B.3) and 
(B.4) that
$${\tilde F}_{g,g^{\prime}}(t)+{\tilde F}_{g,g^{\prime}}(t^{\prime})=
{\tilde F}_{g,g^{\prime}}(t+t^{\prime}) \ {\forall} \ t,t^{\prime}{\in}
{\bf R}_{+}\eqno(B.5)$$
and hence that
$${\tilde F}_{g,g^{\prime}}(nt)=n{\tilde F}_{g,g^{\prime}}(t), \ 
{\forall} \ t{\in}{\bf R}_{+}, \ n{\in}{\bf N}$$
or equivalently
$${\tilde F}_{g,g^{\prime}}(t)=n^{\prime}
{\tilde F}_{g,g^{\prime}}(t/n^{\prime}), \ 
{\forall} \ t{\in}{\bf R}_{+}, \ n^{\prime}{\in}{\bf N}{\backslash}
{\lbrace}0{\rbrace}.$$
These last two equations imply that
$${\tilde F}_{g,g^{\prime}}(rt)=r{\tilde F}_{g,g^{\prime}}(t)$$
for all non-negative $t$ and positive rational $r$; and further, by 
condition ${\bf ({\cal C})}$, this result extends to all positive $r$. 
Hence the action of ${\tilde F}_{g,g^{\prime}}$ on ${\bf R}_{+}$ takes 
the form
$${\tilde F}_{g,g^{\prime}}(t)={\Gamma}(g,g^{\prime})t \ {\forall} \ 
t{\in}{\bf R}_{+},\eqno(B.6)$$
where ${\Gamma}(g,g^{\prime}):={\tilde F}_{g,g^{\prime}}(1)$. By Eqs. 
(B.2) and (B.4), Eq. (B.6) is equivalent to the required formula (B.1); 
and, moreover, it follows from condition {\bf (C.2)} and the continuity 
and linearity of the l.h.s. of that equation with respect to the test 
functions $g$ and $g^{\prime}$ that ${\Gamma}$ is indeed an element of 
${\cal D}_{V}^{{\prime}m}{\otimes}{\cal D}_{V}^{{\prime}m}$ with 
support in the domain 
${\lbrace}(x,x^{\prime}){\in}{\Omega}^{2}{\vert}x^{\prime}=x{\rbrace}$. 
\vskip 0.5cm

\centerline {\bf Appendix C: Proof of Proposition 5.3.}
\vskip 0.3cm
We base the proof of Prop. 5.2 on the following lemma.
\vskip 0.3cm
{\bf Lemma C.1} {\it Let ${\Omega}_{1}$ be any open subset of 
${\Omega}$ whose boundary, ${\partial}{\Omega}_{1}$, does not intersect 
${\partial}{\Omega}$. Then, under the assumptions of Prop. 5.2, the 
restriction of the two-point function ${\Gamma}$ to the 
spatial domain ${\Omega}_{1}^{2}$ is given by a finite sum of the 
following form.
$${\Gamma}(g,g^{\prime})=
{\sum}_{n,n^{\prime}{\in}{\bf N}^{d}}{\sum}_{k,l=1}^{m}
{\sum}_{{\mu},{\nu}=1}^{d}
\int_{\Omega}dxC_{k,l;{\mu},{\nu}}^{n,n^{\prime}}(x)
{\partial}_{x}^{n}g_{k,{\mu}}(x)
{\partial}_{x}^{n^{\prime}}g_{l,{\nu}}^{\prime}(x)$$
$${\forall} \ g,g^{\prime}{\in}{\cal D}_{V}^{m}({\Omega}_{1}), \ 
t,s,t^{\prime},s^{\prime}{\in}{\bf R},\eqno(C.1)$$
where 
\vskip 0.2cm\noindent 
(i) the $C$\rq s are continuous functions on ${\Omega}$ with support in 
some arbitrary neighbourhood of ${\Omega}_{1}$;
\vskip 0.2cm\noindent
(ii) $g_{k,{\mu}}$ is the ${\mu}$\rq th spatial component of the $k$\rq 
th component of $g=(g_{1},. \ ,g_{m})$; and 
\vskip 0.2cm\noindent
(iii) for $n=(n_{1},. \ .,n_{d}){\in}{\bf N}^{d}, \  
{\partial}_{x}^{n}:=
{\partial}^{n_{1}+ \ .+n_{d}}/{\partial}x_{1}^{n_{1}}.. \ 
{\partial}x_{d}^{n_{d}}$.}
\vskip 0.3cm
{\bf Proof of Prop. 5.3 assuming Lemma C.1.} We start by inferring from 
Eq. (5.18) that, for any $g,g^{\prime}{\in}
{\cal D}_{V}^{m}({\Omega}), \ x_{0}{\in}{\Omega}$ and ${\epsilon}$ 
sufficiently small, one can find an open subset ${\Omega}_{1}$ of 
${\Omega}$ such that 
$g_{x_{0},{\epsilon}}$ and $g_{x_{0},{\epsilon}}^{\prime}$ lie in 
${\cal D}_{V}^{m}({\Omega}_{1})$. Hence, by Eqs. (5.18) and (C.1),
$${\Gamma}(g_{x_{0},{\epsilon}},g_{x_{0},{\epsilon}}^{\prime})=$$
$${\sum}_{n,n^{\prime}{\in}{\bf N}^{d}}
{\sum}_{k,l=1}^{m}{\sum}_{{\mu},{\nu}=1}^{d}
{\epsilon}^{-({\vert}n+n^{\prime}{\vert})}
\int_{X}dxC_{k,l;{\mu},{\nu}}^{n,n^{\prime}}(x_{0}+{\epsilon}x)
{\partial}_{x}^{n}g_{k,{\mu}}(x)
{\partial}_{x}^{n^{\prime}}g_{l,{\nu}}^{\prime}(x)$$
$${\forall} \ g,g^{\prime}{\in}{\cal D}_{V}^{m}({\Omega}).\eqno(C.2)$$
where 
${\vert}n+n^{\prime}{\vert}:={\sum}_{k=1}^{d}(n_{k}+n_{k}^{\prime})$: 
evidently the effective domain of integration here is 
${\rm supp}(g){\cap}{\rm supp}(g^{\prime})$. Since the functions $C$ 
are continuous, the summand on the r.h.s. of this equation will 
diverge, as ${\epsilon}{\rightarrow}0$, 
unless {\it either} $n$ and $n^{\prime}$ are both zero {\it or} 
$C_{k,l;{\mu},{\nu}}^{n,n^{\prime}}(x_{0})=0$. Hence the local 
equilibrium condition (5.21) implies that the only non-vanishing $C$\rq 
s are those for which $n$ and $n^{\prime}$ are zero. Thus, Eq. (C.1) 
reduces to the form
$${\Gamma}(g,g^{\prime})={\sum}_{k,l=1}^{m}
{\sum}_{{\mu},{\nu}=1}^{d}
\int_{\Omega}dxC_{k,l;{\mu},{\nu}}^{0,0}(x)g_{k,{\mu}}(x)
g_{l,{\nu}}^{\prime}(x)
{\forall} \ g,g^{\prime}{\in}{\cal D}_{V}^{m}({\Omega}_{1}).
\eqno(C.3)$$
Correspondingly, Eq. (C.2) reduces to the form
$${\Gamma}(g_{x_{0},{\epsilon}},g_{x_{0},{\epsilon}}^{\prime})=
{\sum}_{k,l=1}^{m}{\sum}_{{\mu},{\nu}=1}^{d}
\int_{X}dxC_{k,l;{\mu},{\nu}}^{0,0}(x_{0}+{\epsilon}x)
g_{k,{\mu}}(x)g_{l,{\nu}}^{\prime}(x)$$
$${\forall} \ g,g^{\prime}{\in}{\cal D}_{V}^{m}({\Omega}).\eqno(C.4)$$
It now follows immediately from this formula and the local equilibrium 
condition (5.21) that
$${\sum}_{k,l=1}^{m}{\sum}_{{\mu},{\nu}=1}^{d}
\int_{\Omega}dxC_{k,l;{\mu},{\nu}}^{0,0}(x_{0})g_{k,{\mu}}(x)
g^{\prime}_{l,{\nu}}(x)
=2\bigl(g,K({\theta}(x_{0}))g^{\prime})_{V}$$
$${\forall} \ x_{0}{\in}{\Omega}, \ 
g,g^{\prime}{\in}{\cal D}_{V}^{m}({\Omega}).$$
Further, in view of Eq. (5.17), this last equation signifies that
$$C_{k,l;{\mu},{\nu}}^{0,0}(x)=2K_{kl}\bigl({\theta}(x)\bigr)
{\delta}_{{\mu}{\nu}}
\eqno(C.5)$$
and consequently that Eq. (C.3) reduces to the required formula (5.22), at least for 
$g,g^{\prime}{\in}{\cal D}_{V}^{m}({\Omega}_{1})$. The extension to all 
$g,g^{\prime}$ in ${\cal D}_{V}^{m}({\Omega})$ is trivial, since for any pair of 
elements of the latter space, one can always choose ${\Omega}_{1}$ to be an open 
subset of that space that contains their supports.
\vskip 0.3cm
{\bf Proof of Lemma C.1.} Since the test functions $g_{k,{\mu}}$ and 
$g_{l,{\nu}}^{\prime}$ in Eq. (C.1) are arbitrary elements of 
${\cal D}({\Omega})$, this lemma reduces to the following one.
\vskip 0.3cm 
\vskip 0.3cm
{\bf Lemma C.2.} {\it Let ${\cal T}$ be a 
${\cal D}^{\prime}({\Omega}^{2})$-class distribution whose support lies 
in the region 
${\lbrace}(x,x^{\prime}){\in}{\Omega}^{2}{\vert}x^{\prime}=x{\rbrace}$  
and let ${\Omega}_{1}$ be an open subset of ${\Omega}$ whose boundary, 
${\partial}{\Omega}_{1}$, does not intersect ${\partial}{\Omega}$. Then 
the restriction of ${\cal T}$ to the domain 
${\lbrace}f{\otimes}f^{\prime}{\vert}f,f^{\prime}{\in}
{\cal D}({\Omega}_{1}){\rbrace}$ is given by a finite sum of the form
$${\cal T}(f{\otimes}f^{\prime})={\sum}_{n,n^{\prime}{\in}{\bf N}^{d}}
\int_{\Omega}dxC^{n,n^{\prime}}(x){\partial}_{x}^{n}f(x)
{\partial}_{x}^{n^{\prime}}f^{\prime}(x) \ {\forall} \ 
f,f^{\prime}{\in}{\cal D}({\Omega}_{1}),\eqno(C.6)$$
where the $C$\rq s are continuous functions on ${\Omega}$ with supports 
in some neighbourhood of ${\Omega}_{1}$.}
\vskip 0.3cm
{\bf Proof of Lemma C.2.} Let ${\sigma}$ be a ${\cal D}({\Omega})$-
class function which takes the value unity in ${\Omega}_{1}$ and whose 
support lies in a compact connected subset, $K$, of ${\Omega}$ whose 
boundary, ${\partial}K$, does not intersect either ${\partial}{\Omega}$ 
or ${\partial}{\Omega}_{1}$. We define the distribution 
${\tilde {\cal T}} \ \bigl({\in}{\cal D}^{\prime}({\Omega}^{2})\bigr)$ 
by the formula
$${\tilde {\cal T}}(x,x^{\prime})={\sigma}(x){\sigma}(x^{\prime})
{\cal T}(x,x^{\prime}).\eqno(C.7)$$
Thus, ${\tilde {\cal T}}$ coincides with ${\cal T}$ in 
${\Omega}_{1}^{2}$ and
$${\rm supp}({\tilde {\cal T}}){\subset}{\lbrace}(x,x^{\prime}){\in}
K^{2}{\vert}x^{\prime}=x{\rbrace}.\eqno(C.8)$$
\vskip 0.2cm
We define ${\Phi}$ to be the linear transformation of $X^{2}$ given by 
the formula
$${\Phi}(y,z)=(y+z,y-z) \ {\forall} \ y,z{\in}X\eqno(C.9)$$
from which its follows that
$${\Phi}^{-1}(x,x^{\prime})=\bigl({1\over 2}(x+x^{\prime}),
{1\over 2}(x-x^{\prime})\bigr) \ {\forall} \ 
x,x^{\prime}{\in}X.\eqno(C.10)$$ 
We then define
$${\Theta}:={\Phi}^{-1}
({\Omega}^{2})={\lbrace}(y,z){\in}X^{2}{\vert}(y{\pm}z){\in}
{\Omega}{\rbrace},$$
and we define the bijection $F{\rightarrow}{\hat F}$ of 
${\cal D}({\Omega}^{2})$ onto ${\cal D}({\Theta})$ by the formula 
${\hat F}=F{\circ}{\Phi}$, i.e.
$${\hat F}(y,z)=F(y+z,y-z) \ {\forall} \ 
(y,z){\in}{\Theta}.\eqno(C.11)$$
Correspondingly we define the distribution ${\hat {\cal T}} \ ({\in}
{\cal D}^{\prime}({\Theta}))$ in terms of ${\tilde {\cal T}}$ by the 
formula
$${\hat {\cal T}}({\hat F})={\tilde {\cal T}}(F) \ {\forall} \ 
F{\in}{\cal D}({\Omega}^{2}).\eqno(C.12)$$
It follows from Eqs. (C.8), (C.11) and (C.12) that
$${\rm supp}({\hat {\cal T}}){\subset}K{\times}{\lbrace}0{\rbrace}.
\eqno(C.13)$$
\vskip 0.2cm
We want to restrict ${\hat {\cal T}}$ to an open subset of ${\Theta}$ 
which contains the support of this distribution and takes the form 
${\Omega}_{2}{\times}J$, where ${\Omega}_{2}$ and $J$ are open subsets 
of ${\Omega}$ and $X$ respectively. Accordingly, we choose $b$ to be a  
positive number that is less than 
$dist({\partial}K,{\partial}{\Omega})$, the minimal distance between 
the boundaries, ${\partial}K$ and ${\partial}{\Omega}$, of $K$ and 
${\Omega}$. We then define ${\Omega}_{2}:=
{\lbrace}y{\in}X{\vert}(y,z){\in}{\Theta} \ {\forall} \ 
{\vert}z{\vert}{\leq}b{\rbrace}$ and 
$J:={\lbrace}z{\in}X{\vert}{\vert}z{\vert}
<b{\rbrace}$. It follows from these definitions that ${\Omega}_{2}$ and 
${\Omega}_{2}{\times}J$ are open subsets of ${\Omega}$ and ${\Theta}$, 
respectively, that $K{\subset}{\Omega}_{2}$ and that 
${\partial}{\Omega}_{2}$, the boundary of ${\Omega}_{2}$, does not 
intersect either ${\partial}K$ or ${\partial}{\Omega}$. Hence, by Eq. 
(C.13), ${\Omega}_{2}{\times}J$ is an open neighbourhood of 
${\rm supp}({\hat {\cal T}})$ and the restriction, 
${\hat {\cal T}}^{\prime}$, of ${\hat {\cal T}}$ to this domain carries 
all the information we require. It follows from its definition that 
${\hat {\cal T}}^{\prime}{\in}{\cal 
D}^{\prime}({\Omega}_{2}{\times}J)$.
\vskip 0.2cm
Now let $e$ be an arbitrary element of ${\cal D}({\Omega}_{2})$. Then 
for $e^{\prime}{\in}{\cal D}(J), \ {\hat {\cal T}}^{\prime}$ induces a 
continuous linear functional ${\hat {\cal T}}_{e}^{\prime}$ on 
${\cal D}(J)$ according to the formula
$${\hat {\cal T}}_{e}^{\prime}(e^{\prime})=
{\hat {\cal T}}^{\prime}(e{\otimes}e^{\prime}) \ 
{\forall} \ e^{\prime}{\in}{\cal D}(J),\eqno(C.14).$$
where the mapping $e{\rightarrow}{\hat {\cal T}}_{e}^{\prime}$ of 
${\cal D}({\Omega}_{2})$ into ${\cal D}^{\prime}(J)$ is continuous. 
Further, it follows from Eqs. (C.13) and (C.14) that 
${\hat {\cal T}}_{e}^{\prime}$ has support at the origin and 
consequently, by Schwartz\rq s point support theorem [33, Theorem 35], 
that this distribution is a finite sum of derivatives of ${\delta}(z)$, 
with coefficients given by linear continuous functionals of $e$, i.e.
$${\hat {\cal T}}_{e}^{\prime}(e^{\prime})={\sum}_{n}T_{n}(e)
({\partial}^{n}e^{\prime})(0),\eqno(C.15)$$
where each $T_{n}{\in}{\cal D}^{\prime}(J)$. Further, in view of the 
definition of ${\hat {\cal T}}_{e}^{\prime}$, it follows from Eqs. 
(C.13) and (C.14) that $T_{n}$ has support in the compact $K$ and 
therefore, by Schwartz\rq s compact support theorem [33, Theorem 26], 
it is a finite sum of derivatives of continuous functions on 
${\Omega}_{2}$ with support in an arbitrary neighbourhood of $K$. 
Consequently, by Eq. (C.15), the action of 
${\hat {\cal T}}^{\prime}$ on 
${\cal D}({\Omega}{\times}J)$ is given by a finite sum of the form 
$${\hat {\cal T}}^{\prime}({\hat F})=
{\sum}_{n^{\prime},n}\int_{{\Omega}_{2}}dy{\hat D}^{n^{\prime},n}(y)
{\partial}_{y}^{n^{\prime}}{\partial}_{z}^{n}
{\hat F}(y,z)_{z=0} \ 
{\forall} \ {\hat F}{\in}{\cal D}({\Omega}_{2}{\times}J),\eqno(C.16)$$
where the ${\hat D}^{\prime}$s are continuous functions on 
${\Omega}_{2}$ with support in a neighbourhood of $K$. Hence, as 
${\hat {\cal T}}^{\prime}$ is just the restriction of ${\hat {\cal T}}$ 
to ${\cal D}({\Omega}_{2}{\times}J)$ and since ${\tilde {\cal T}}$ 
coincides with ${\cal T}$ in ${\Omega}_{1}^{2}$, it follows from Eqs. 
(C.9)-(C.12) that Eq. (C.16) is equivalent to the formula
$${\cal T}(F)={\sum}_{n,n^{\prime}{\in}{\bf N}^{d}}
\int_{\Omega}dxC^{n,n^{\prime}}(x){\partial}_{x}^{n}
{\partial}_{x^{\prime}}^{n^{\prime}}
F(x,x^{\prime})_{{\vert}x^{\prime}=x} \ {\forall} \ 
f,f^{\prime}{\in}{\cal D}({\Omega}_{1}),\eqno(C.17)$$
which in turn is equivalent to the required Eq. (C.6).
\vskip 0.5cm
\centerline {\bf Appendix D: Proof of Proposition 5.5}
\vskip 0.3cm
{\bf Part (a).} The characteristic functional for the process ${\xi}$ 
is 
$$C(f^{(1)},. \ .,f^{(r)};t_{1},. \ .,t_{r})=E\bigl[{\rm exp}
\bigl(i{\sum}_{k=1}^{r}{\xi}_{t_{k}}(f^{(k)})\bigr)\bigr]$$ 
$${\forall} \ f^{(1)},. \ .,f^{(r)}{\in}{\cal D}^{m}({\Omega});
\ t_{1},. \ .,t_{r}{\in}{\bf R}, \ r{\in}{\bf N}.
\eqno(D.1)$$
Equivalently, since the process is stationary, 
$$C(f^{(1)},. \ .,f^{(r)};t_{1},. \ .,t_{r})=E\bigl[{\rm exp}
\bigl(i{\sum}_{k=1}^{r}{\xi}_{t_{k}+t_{0}}(f^{(k)})\bigr)\bigr]$$ 
$${\forall} \ f^{(1)},. \ .,f^{(r)}{\in}{\cal D}^{m}({\Omega});
\ t_{0},t_{1},. \ .,t_{r}{\in}{\bf R}, \ r{\in}{\bf N}.\eqno(D.2)$$
Here we are at liberty to choose $t_{0}$ to be any real number and, for 
any specified set of times $t_{1},. \ .,t_{r}$, we choose it so that 
$t_{1}+t_{0},. \ ,t_{r}+t_{0}$ are all positive. It then follows from 
Eq. (5.28) that
$${\xi}_{t_{k}+t_{0}}(f^{(k)})={\xi}(T_{t_{k}+t_{0}}^{\star}f^{(k)})+
\int_{0}^{t_{k}+t_{0}}dw_{u,0}(T_{t_{k}+t_{0}-u}^{\star}f^{(k)})$$
and therefore that Eq. (D.2) may be re-expressed as 
$$C(f^{(1)},. \ .,f^{(r)};t_{1},. \ .,t_{r})=$$
$$E\bigl[{\rm exp}
\bigl(i{\sum}_{k=1}^{r}{\xi}(T_{t_{k}+t_{0}}^{\star}f^{(k)})\bigr)
{\rm exp}\bigl(i{\sum}_{k=1}^{r}\int_{0}^{t_{k}+t_{0}}
dw_{u,0}(T_{t_{k}+t_{0}-u}^{\star}f^{(k)})\bigr)\bigr].\eqno(D.3)$$
We now define
$${\tilde C}(f^{(1)},. \ .,f^{(r)};t_{0},t_{1},. \ .,t_{r})=
{\rm exp}\bigl(i{\sum}_{k=1}^{r}\int_{0}^{t_{k}+t_{0}}dw_{u,0}
(T_{t_{k}+t_{0}-u}^{\star}f^{(k)})\bigr)\eqno(D.4)$$
and, using the Schwartz inequality, we infer from the last two
equations that
$${\vert}C(f^{(1)},. \ .,f^{(r)};t_{1},. \ .,t_{r})-
{\tilde C}(f^{(1)},. \ .,f^{(r)};t_{0},t_{1},. \
.,t_{r}){\vert}^{2}$$
$${\leq}E\bigl[{\vert}{\rm exp}
\bigl(i{\sum}_{k=1}^{r}{\xi}
(T_{t_{k}+t_{0}}^{\star}f^{(k)})\bigr)-1{\vert}^{2}\bigr]$$
$${\leq}E\bigl[\big({\sum}_{k=1}^{r}{\xi}
(T_{t_{k}+t_{0}}^{\star}f^{(k)})\bigr)^{2}\bigr]=
{\sum}_{k,l=1}^{r}E\bigl({\xi}
(T_{t_{k}+t_{0}}^{\star}f^{(k)})
{\xi}(T_{t_{l}+t_{0}}^{\star}f^{(l)})\bigr).$$
It follows from the dissipativity condition (3.12) that the r.h.s.
of this estimate vanishes in the limit $t_{0}{\rightarrow}{\infty}$,
and therefore that
$$C(f^{(1)},. \ .,f^{(r)};t_{1},. \ .,t_{r})=
{\rm lim}_{t_{0}{\rightarrow}{\infty}}
{\tilde C}(f^{(1)},. \ .,f^{(r)};t_{0},t_{1},. \ .,t_{r}),$$
i.e., by Eq. (D.4), that 
$$C(f^{(1)},. \ .,f^{(r)};t_{1},. \ .,t_{r})=
{\rm lim}_{t_{0}{\rightarrow}{\infty}}
E\bigl[{\rm exp}\bigl(i{\sum}_{k=1}^{r}
\int_{0}^{t_{k}+t_{0}}dw_{u,0}(T_{t_{k}+t_{0}-
u}^{\star}f^{(k)})\bigr)\bigr].$$
Since, by Eq. (5.7) and the chaoticity condition {\bf (C.1)}, the
process $w$ is Gaussian, it follows immediately from this
last equation that the process ${\xi}$ is Gaussian.
\vskip 0.2cm
In order to show that it is also Markovian, we need just to
prove that, for $t{\in}{\bf R}$ and any random variable
$B_{{\geq}t}$ generated by
${\lbrace}{\xi}_{u}(f){\vert}f{\in}{\cal D}^{m}({\Omega}), \
u{\geq}t{\rbrace}$, the conditional expectations of $B_{{\geq}t}$ with 
respect to the random variables for time $t$ and for times ${\leq}t$ 
are equal, i.e. that
$$E(B_{{\geq}t}{\vert}{\xi}_{t})= 
E(B_{{\geq}t}{\vert}{\xi}_{{\leq}t}).\eqno(D.5)$$
Now the random variables over the times ${\geq},=$ and ${\leq}t$
are generated by linear combinations of terms
$$F_{{\geq}t}={\rm exp}
\bigl(i{\sum}_{k=1}^{p}{\xi}_{u_{k}}(f^{(k)})\bigr),\eqno(D.6)$$
$$F_{t}={\rm exp}\bigl(i{\xi}_{t}(f)\bigr)\eqno(D.7)$$
and
$$F_{{\leq}t}={\rm exp}
\bigl(i{\sum}_{l=1}^{r}{\xi}_{s_{l}}(f^{{\prime}(l)})\bigr),
\eqno(D.8)$$
respectively, where $u_{k}{\geq}t{\geq}s_{l}$ and $f^{(k)}, \ f$
and $f^{{\prime}(l)}$ are elements of ${\cal D}^{m}({\Omega})$.
\vskip 0.2cm
It follows from Eqs. (D.1) and (D.6)-(D.8), together with the
Gaussian property of ${\xi}$, that
$$E(F_{{\geq}t}F_{t})=
C(f^{(1)},. \ .,f^{(p)};u_{1},. \ .,u_{p})C(f;t)
{\rm exp}\bigl[-{\sum}_{k=1}^{p}E\bigl({\xi}_{u_{k}}(f^{(k)})
{\xi}_{t}(f)\bigr)\bigr]\eqno(D.9)$$
and that
$$E(F_{{\geq}t}F_{{\leq}t})=
C(f^{(1)},. \ .,f^{(p)};u_{1},. \ .,u_{p})
C(f^{{\prime}(1)},. \ .,f^{{\prime}(r)};s_{1},. \
.,s_{r}){\times}$$
$${\rm exp}\bigl[-{\sum}_{k=1}^{p}{\sum}_{l=1}^{r}
E\bigl({\xi}_{u_{k}}(f^{(k)})
{\xi}_{s_{l}}(f^{{\prime}(l)})\bigr)\bigr].
\eqno(D.10)$$
Further, since $u_{k}{\geq}t{\geq}s_{l}$, it follows from Eqs.
(4.5) and (4.6) that the summands appearing in the exponents 
in Eqs. (D.9) and (D.10) are equal to 
$E\bigl({\xi}(T_{u_{k}-t}^{\star}f^{(k)}){\xi}(f)\bigr)$
and $E\bigl({\xi}(T_{u_{k}-s_{l}}^{\star}f^{(k)})
{\xi}(f^{{\prime}(l)})\bigr)$, respectively, and
therefore those equations may be re-expressed as
$$E(F_{{\geq}t}F_{t})=
C(f^{(1)},. \ .,f^{(p)};u_{1},. \ .,u_{p})C(f;t)
{\rm exp}\bigl[-{\sum}_{k=1}^{p}E\bigl({\xi}
(T_{u_{k}-t}^{\star}f^{(k)}){\xi}(f)\bigr)\bigr]
\eqno(D.11)$$
and
$$E(F_{{\geq}t}F_{{\leq}t})=
C(f^{(1)},. \ .,f^{(p)};u_{1},. \ .,u_{p})
C(f^{{\prime}(1)},. \ .,f^{{\prime}®};s_{1},. \
.,s_{r}){\times}$$
$${\rm exp}\bigl[-{\sum}_{k=1}^{p}{\sum}_{l=1}^{r}
E\bigl({\xi}(T_{u_{k}-s_{l}}^{\star}f^{(k)})
{\xi}(f^{{\prime}(l)})\bigr)\bigr].\eqno(D.12)$$
Further, since $E(F_{{\geq}t}{\vert}{\xi}_{t})$ is the unique random 
variable of the ${\xi}$-process at time $t$ for which
$$E\bigl(E(F_{{\geq}t}{\vert}{\xi}_{t})F_{t}\bigr)=
E(F_{{\geq}t}F_{t})$$
for all $F_{{\geq}t}$ and $F_{t}$ of the forms given by Eqs. (D.6) and 
(D.7), respectively, it follows from Eq. (D.9), together with the 
stationarity and the Gaussian property of the process, that
$$E(F_{{\geq}t}{\vert}{\xi}_{t})=
{C(f^{(1)},. \ .,f^{(p)};u_{1},. \ .,u_{p})\over 
C(T_{u_{1}-t}^{\star}f^{(1)},. \ .,T_{u_{p}-t}^{\star}f^{(p)};
0,. \ .,0)}{\rm exp}\bigl(i{\sum}_{k=1}^{p}{\xi}_{t}
(T_{u_{k}-t}^{\star}f^{(k)}\bigr).\eqno(D.13)$$ 
Hence, by Eq. (D.8), 
$$E\bigl(E(F_{{\geq}t}{\vert}{\xi}_{t})F_{{\leq}t}\bigr)=$$
$${C(f^{(1)},. \ .,f^{(p)};u_{1},. \ .,u_{p})\over 
C(T_{u_{1}-t}^{\star}f^{(1)},. \ .,T_{u_{p}-t}^{\star}f^{(p)};
0,. \ .,0)}C\bigl(f^{{\prime}(1)},. \ 
.,f^{{\prime}(r)},{\sum}_{k=1}^{p}
T_{u_{k}-t}^{\star}f^{(k)}:s_{1},. \ .,s_{r},t\bigr).\eqno(D.14)$$
Further, in view of the Gaussian property of the process, the last 
factor in this formula is equal to
$$C(f^{{\prime}(1)},. \ .,f^{{\prime}(r)}:s_{1},. \ .,s_{r})
C({\sum}_{k=1}^{p}T_{u_{k}-t}^{\star}f^{(k)};t){\times}$$ 
$${\rm exp}\bigl[-{\sum}_{k=1}^{p}{\sum}_{l=1}^{r}
E\bigl({\xi}_{t}(T_{u_{k}-t}^{\star}f^{(k)}{\xi}_{s_{l}}
(f^{{\prime}(l)})\bigr)\bigr].$$
Therefore since, by Eq. (4.6) and the semigroup property of 
$T^{\star}({\bf R}_{+})$, the summand in the exponent in this 
expression is equal to $E\bigl({\xi}(T_{u_{k}-s_{l}}^{\star}f^{(k)}
{\xi}(f^{{\prime}(l)})\bigr)$, it follows from Eqs. (D.10) and (D.14) 
that
$$E\bigl(E(F_{{\geq}t}{\vert}{\xi}_{t})F_{{\leq}t}\bigr)=
E(F_{{\geq}t}F_{{\leq}t}).$$
Hence
$$E(F_{{\geq}t}{\vert}{\xi}_{t})=
E(F_{{\geq}t}{\vert}{\xi}_{{\leq}t}),$$
which signifies that the process is temporally Markovian.
\vskip 0.3cm 
{\bf Part (b).} Since Eq. (5.8) implies that $w_{t,s}=-w_{s,t}$ and 
since $w_{t,s}$ and ${\xi}_{u}$ are Gaussian random fields whose means 
are zero, it follows from Eq. (5.9) that the latter two fields are 
statistically independent of one another if $s$ and $t$ are both 
greater than or equal to $u$.
\vfill\eject
\centerline {\bf References}
\vskip 0.3cm\noindent
1. R. Graham and H. Haken, Z. Phys. {\bf 237} 31, 1970
\vskip 0.2cm\noindent
2. K. Hepp and E. H. Lieb, Helv. Phys. Acta  {\bf 46} (1973), 573. 
\vskip 0.2cm\noindent
3. G. Alli and G. L. Sewell, J. Math. Phys. {\bf 36} (1995), 5598.
\vskip 0.2cm\noindent
4. F. Bagarello and G. L. Sewell, J. Math. Phys. {\bf 39}, (1998), 
2730.
\vskip 0.2cm\noindent
5. P. Glansdorf and I. Prigogine, Thermodynamic Theory of Structure, 
Stability and Fluctuations, Wiley-Interscience, London, 1971.
\vskip 0.2cm\noindent
6. H. Fr${\rm {\ddot o}}$hlich, Int. J. Quantum Chem. {\bf 2} (1968), 
641.
\vskip 0.2cm\noindent
7. G. Gallavotti, J. Stat. Phys. {\bf 84} (1996), 899.
\vskip 0.2cm\noindent
8. D. Ruelle, J. Stat. Phys. {\bf 85} (1996), 1.
\vskip 0.2cm\noindent
9. H. Spohn, J. Phys. A {\bf 16} (1983), 4275.
\vskip 0.2cm\noindent
10. B. Derrida, J. L. Lebowitz amd E. R. Speer, 
J. Stat. Phys. {\bf 107} (2002), 599.
\vskip 0.2cm\noindent
11. L. Bertini, A. de Sole, D. Gabrielli, G. Jona-Lasinio and
C. Landim, J. Stat. Phys. {\bf 107} (2002), 635.
\vskip 0.2cm\noindent
12. D. Ruelle, J. Stat. Phys. {\bf 98} (2000), 57.
\vskip 0.2cm\noindent
13. S. Tasaki and T. Matsui, Fluctuation theorem, nonequilibrium steady 
states and the McLennan-Zubarev ensembles of $L^{1}$-asymptotical 
abelian $C^{\star}$-dynamical systems, in Fundamental aspects of 
quantum physics, Ed. L. Accardi and S. Tasaki, World Scientific, 
Singapore, 2003, Pp. 100-119.
\vskip 0.2cm\noindent
14. G. L. Sewell, Lett. Math. Phys. {\bf 68} (2004), 53. 
\vskip 0.2cm\noindent
15. G. L. Sewell, Quantum Mechanics and its Emergent
Macrophysics, Princeton Univ. Press, Princeton, Oxford, 2002.
\vskip 0.2cm\noindent
16. G. L. Sewell, Quantum macrostatistics and irreversible 
thermodynamics, in
of Lec. Notes in Mathematics, Vol. 1442, Ed. L. Accardi and W. Von 
Waldebfels, Springer, Berlin, 1990, Pp.368-83.
\vskip 0.2cm\noindent
17. L. Onsager, Phys. Rev. {\bf 37} (1931), 405; {\bf 38} (1931), 2265  
\vskip 0.2cm\noindent
18. L. D. Landau and E. M. Lifschitz, Fluid Mechanics,
Pergamon, Oxford, 1984.
\vskip 0.2cm\noindent
19. L. Boltzmann, Lectures on Gas Theory, University of California 
Press, 
Berkeley, CA, 1964.
\vskip 0.2cm\noindent
20. O. E. Lanford, Time evolution of large classical systems, in 1974 
Battelle Rencontre, Ed. J.Moser, LNP 38, Springer, Berlin, 1975.
\vskip 0.2cm\noindent
21. T. G. Ho, L. J. Landau and A. J. Wilkins, On the weak coupling 
limit for a Fermi gas in a random potential, Rev. Math. Phys. {\bf 5} 
(1993), 209. 
\vskip 0.2cm\noindent
22. L. Onsager and S. Machlup, Phys. Rev. {\bf 91} (1953), 1505.
\vskip 0.2cm\noindent
23. G. Grinstein, D.-H. Lee and S. Sachdev, Phys. Rev. Lett. 
{\bf 64} (1990), 1927.
\vskip 0.2cm\noindent
24. J. R. Dorfman, T. R. Kirkpatrick and J. V. Sengers, Annu.
Rev. Phys. Chem. {\bf 45} (1994), 213.
\vskip 0.2cm\noindent
25. R. F. Streater and A. S. Wightman, PCT, Spin and
Statistics, and All That, W. A. Benjamin, New York, 1964.
\vskip 0.2cm\noindent
26. R. Haag, N. M. Hugenholtz and M. Winnink, Commun. Math. Phys. {\bf 
5}, (1967), 215.
\vskip 0.2cm\noindent
27. D. Ruelle, Statistical Mechanics, W. A. Benjamin, New York, 1969.
\vskip 0.2cm\noindent
28. G. G. Emch, Algebraic Methods in Statistical Mechanics and Quantum 
Field Theory, Wiley, New York, 1971.
\vskip 0.2cm\noindent
29. G. G. Emch, H. J. F. Knops and E. J. Verboven, J. Math. Phys. {\bf 
11} (1970), 1655.
\vskip 0.2cm\noindent
30. V. Jakcic and C. A. Pillet, Commun. Math. Phys. {\bf 226} (2002), 
131.
\vskip 0.2cm\noindent
31. I. E. Segal, Ann. Math. {\bf 48} (1947), 930.
\vskip 0.2cm\noindent
32. G. L. Sewell, J. Math. Phys. {\bf 11} (1970), 1868. 
\vskip 0.2cm\noindent
33. L. Schwartz: Th\'eorie des Distributions, 
Hermann, Paris, 1998. 
\vskip 0.2cm\noindent
34. L. Accardi, A. Frigerio and J. T. Lewis, Publ. RIMS {\bf 18} 
(1982), 97.  
\vskip 0.2cm\noindent
35. E. Nelson, Ann. Math. {\bf 70} (1959), 572.
\vskip 0.2cm\noindent
36. E. Nelson, Dynamical theories of Brownian motion,
Princeton Univ. Press, Princeton, 1972.
\vskip 0.2cm\noindent
37. R. Sen and G. L. Sewell, J. Math. Phys. {\bf 43} (2002), 1323.
\vskip 0.2cm\noindent
38. B. Nachtergaele and H. T. Yau, Commun. Math. Phys. {\bf 243} 
(2003), 485.
\vskip 0.2cm\noindent
39. D. Goderis, P. Vets and A. Verbeure, Prob. Th. Re. Fields
{\bf 82} (1989), 527. 
\vskip 0.2cm\noindent
40. M. Broidio, B. Momont and A. Verbeure, J. Math. Phys. {\bf 36} 
(1995), 6746.
\vskip 0.2cm\noindent
41. H. G. B. Casimir, Rev. Mod. Phys. {\bf 17} (1945), 343-50.
\end